%% file: dyn5.tex
\documentclass[11pt]{article}

\newcommand{\be}{\begin{equation}}
\newcommand{\ee}{\end{equation}}
\newcommand{\ben}{\begin{eqnarray}}
\newcommand{\een}{\end{eqnarray}}
\newcommand{\bR}{\mbox{\boldmath $R$}}

\newcommand{\bi}{\mbox{\boldmath $i$}}

\newcommand{\bF}{\mbox{\boldmath $F$}}

\newcommand{\br}{\mbox{\boldmath $r$}}

\newcommand{\bv}{\mbox{\boldmath $v$}}
\newcommand{\bT}{\mbox{\boldmath $T$}}

\newcommand{\bp}{\mbox{\boldmath $p$}}
\newcommand{\bA}{\mbox{\boldmath $A$}}

\usepackage{setspace}
\usepackage{graphicx}
\usepackage{epstopdf}
\usepackage{dcolumn}
\usepackage{bm}
\usepackage{cite}
\usepackage{verbatim}


\begin{document}
\title{Asteroid Deflection Using a Spacecraft in Restricted Keplerian Motion}
\author{Yohannes Ketema\\Dept. of Aerospace Engineering and Mechanics\\University of Minnesota, Minneapolis, MN 55455}
\maketitle

\begin{abstract}
A method for asteroid deflection that makes use of a spacecraft moving
back and forth on a segment of a Keplerian orbit about the asteroid is
described and studied. It is shown that on average the spacecraft 
can exert a significantly larger force on the asteroid than e.g. a
stationary gravity   tractor, thereby reducing the time needed to
effect the desired deflection of the asteroid. Furthermore, the current method does not require
canted thrusters on the spacecraft (unlike a stationary gravity tractor) markedly reducing the amount of fuel
needed for a given deflection to be realized. The
method also allows for the simultaneous use of several spacecraft,
further strengthening the overall tugging effect on the asteroid, and distributing the thrust requirement among the spacecraft.
\end{abstract}
\section*{Nomenclature}
\begin{tabular}{ l c l}
 $m_a$&=&mass of asteroid \\
$m_c$&=&mass of spacecraft\\
$\mu_a$&=&gravitational parameter for asteroid\\
$\mu_c$&=&gravitational parameter for spacecraft\\
$r_a$&=&radius of asteroid\\
$r$&=&distance to spacecraft from asteroid center \\
$\alpha$&=&ratio of spacecraft distance to asteroid radius\\
$G$&=&universal gravitational constant\\
$\theta$&=&true anomaly on spacecraft orbit about asteroid\\
$\theta_b$&=&bounding angle on spacecraft orbit about asteroid \\
$\Delta v$&=&impulsive velocity change\\
$I$&=&impulse imparted to asteroid \\
$h$&=&specific angular momentum for spacecraft orbit about asteroid\\
$r_p$&=&periapsis distance for spacecraft orbit about asteroid\\
$r_{pm}$&=&smallest allowable periapsis distance for spacecraft orbit about asteroid\\
$e$&=&eccentricity of spacecraft orbit about asteroid\\
$\gamma$&=&flight path angle for spacecraft orbit about asteroid\\
$\varphi$&=&plume half angle\\
$v_{esc}$&=&escape velocity for motion with respect to asteroid\\
$t_m$&=&time of flight on spacecraft orbit segment \\
\end{tabular}

\noindent
\begin{tabular}{ l c l}
$\eta_k(\theta_b,\varphi,e)$&=&nondimensional average force from spacecraft in orbit\\
$\eta_k(\theta_b,\varphi)$&=&nondimensional average force from spacecraft in circular orbit\\
$\eta_s$&=&nondimensional average force from stationary gravity tractor\\
$\beta$&=&angle between asteroid-spacecraft vector and line from
  spacecraft tangent to the asteroid surface\\
$f_{do}$&=&force exerted on asteroid by spacecraft in displaced
  orbit\\
$\zeta_k(\theta_b,e)$&=&nondimensional impulse per unit mass of fuel used in
  orbiting spacecraft\\
$\zeta_s$&=&nondimensional impulse per unit mass of fuel used in
  stationary spacecraft\\
$\nu(\theta_b,e,\varphi)$&=&nondimensional impulsive velocity change\\
\end{tabular}

\section{Introduction}

The potential for Earth-asteroid impacts \cite{chapman,binzel} has recently generated
much interest in the development of techniques to avert a projected
collision by modifying the orbit of the asteroid. 
A broad study of possible
approaches to the deflection of asteroids is presented in \cite{sanchez} where
various methods are compared, and an assessment is made of their
technology readiness. Asteroid deflection schemes analyzed include
nuclear interceptors, kinetic impactors, and gravity tractors. 

Nuclear interceptors are meant to detonate a nuclear
charge on or at a distance from  
the surface of an asteroid \cite{sanchez,
  ahrens,dearborn,hanrahan}. When considering asteroid deflection
using a nuclear charge,
it is important to take into account
the response of the asteroid body to the nuclear detonation, i.e. whether it might break up
into small fragments, or if the asteroid will remain a single body, which
would be the ideal case. Therefore,
the material composition of the asteroid and its structural
characteristics are important factors that 
must be studied. It has been shown \cite{sanchez} that the distance from the surface of the
asteroid to the detonating nuclear charge can be used as a parameter
to effect the latter scenario.

Kinetic impactors aim to impart an impulse to the asteroid through a direct collision. Studies of  kinetic impactors can be found in
\cite{ahrens,mcinnes,dachwald,izzo1,asfaw}. As in the case of nuclear interceptors, careful consideration must be given to the
composition of the asteroid to ensure that it does not break up during the collision. 

A method of asteroid deflection that is unaffected by the composition of the
asteroid is the gravity-tractor, where a spacecraft exerts a force on the asteroid purely through 
gravitational coupling.  Since the gravity tractor was first proposed in \cite{lu-love} it has been the 
subject of several studies on topics such as the optimal control of
its position with respect to the asteroid, and realistic studies of implementation for the deflection of specific asteroids
\cite{gehler,wei,olympio,fahn}.

In its basic form (as will be detailed further in the body of this paper) the gravity tractor uses
a pair of thrusters that are canted with respect to the direction of
the force to be exerted on the asteroid. This is necessary so that exhaust plumes do not impinge on the
asteroid directly, as this would counteract  the gravitational
tugging effect. However, the canting of the thrusters means that  a considerable
amount of thrust goes unused in terms of asteroid deflection, making the gravity tractor
rather inefficient with respect to fuel use.

A gravity tractor that does not necessarily require canted thrusters
has been suggested in \cite{mcinnes2} 
where the spacecraft is in a displaced non-Keplerian orbit
with respect to the asteroid. It is shown that when the offset distance of the orbit
from the center of the asteroid is sufficiently large (to avoid plume
impingement), this method can produce the same magnitude of net force on the asteroid as the stationary gravity tractor, for a 25\% less
thrust developed by the spacecraft's thruster. The requirement of
relatively large offset distances, however, means that the force from
the displaced-orbit gravity tractor is generally small -- in
the same order of 
magnitude as that of the stationary gravity tractor. Also, for the
values of the offset required to avoid plume impingement, the
displaced orbit is unstable as follows from the analysis in
\cite{mcinnes3}, and must be stabilized using state feedback.

The use of formation flying spacecraft
in displaced non-Keplerian orbits with solar sails, and associated
problems of controlling the formation are studied in
\cite{gong}. This approach, while similar to the
displaced-orbits-method in
\cite{mcinnes2} removes the problem of 
plume impingement.

A common characteristic of the above mentioned approaches to asteroid
deflection is that the perturbing forces applied on the asteroid are
generally small (compared to the force of gravity from the Sun on the
asteroid). However, as shown in \cite{izzo}, the effect of these 
perturbing forces on the resulting deflection increases with passing time through a secular
effect. Thus it is advantageous to start the deflection action as
early as possible before the projected time of collision.
On the other hand, the amount of time available between the detection
of an asteroid and the projected time of collision may not be
large. It is therefore important to study ways of increasing
the forces that are exerted on the asteroid, regardless of what
specific method is being used.

The main goal of the  current paper is therefore to present 
a gravity tractor that on average can exert a significantly larger force
on the asteroid than the gravity-based methods mentioned above, for the
same spacecraft mass.
In this method the spacecraft describes what will be referred
to as {\em restricted Keplerian
motion} about the asteroid. The motion, as will be described
further in the body of this paper, consists of the spacecraft moving back and
forth along a segment of a Keplerian orbit, changing the direction of
its velocity at the ends of the segment through impulsive thrusts. In
addition
to the larger average force exerted on the asteroid
when using the current method, the net impulse on the
asteroid per mass of fuel used in the spacecraft is significantly
larger than in the case of the stationary gravity tractor. This
translates into higher efficiency in terms of deflection effected per
amount of fuel used. The advantages gained in a larger average force and increased impulse
per mass, compared to other gravitationally based methods,  derive
from the fact that the spacecraft can generally come
closer to the surface of the asteroid while saving fuel by not
having to use canted thrusters. 

The rest of this paper is organized as follows: In Section 2 the
forces on the asteroid-spacecraft system are identified and the
equations of motion for the asteroid and for the spacecraft are
written. In Section 3 an
expression for the deflection of the asteroid is derived in terms of the
characteristics of the restricted Keplerian motion. Next in
Section 4, the force exerted by the gravity tractor on the asteroid is studied and
compared to the corresponding forces from a  stationary gravity
tractor and a displaced-orbit gravity tractor for a given spacecraft mass.
The performance of the gravity tractor from the point of view of fuel
expenditure is studied in Section 5, where
a measure of fuel
efficiency is defined and compared between the current gravity
tractor, the stationary gravity tractor, and the displaced-orbit
gravity tractor. A numerical example is presented in Section 6 where a
hypothetical deflection of Asteroid 2007 VK184 is studied and compared
to existing results using other methods. Lastly, concluding remarks are given in
Section 7.

\section{Forces on the asteroid}

\begin{figure}[htbp]
  \begin{center}
    \unitlength=.5in
    \begin{picture}(4,6)
     \put(0,0){\includegraphics[scale=0.8]{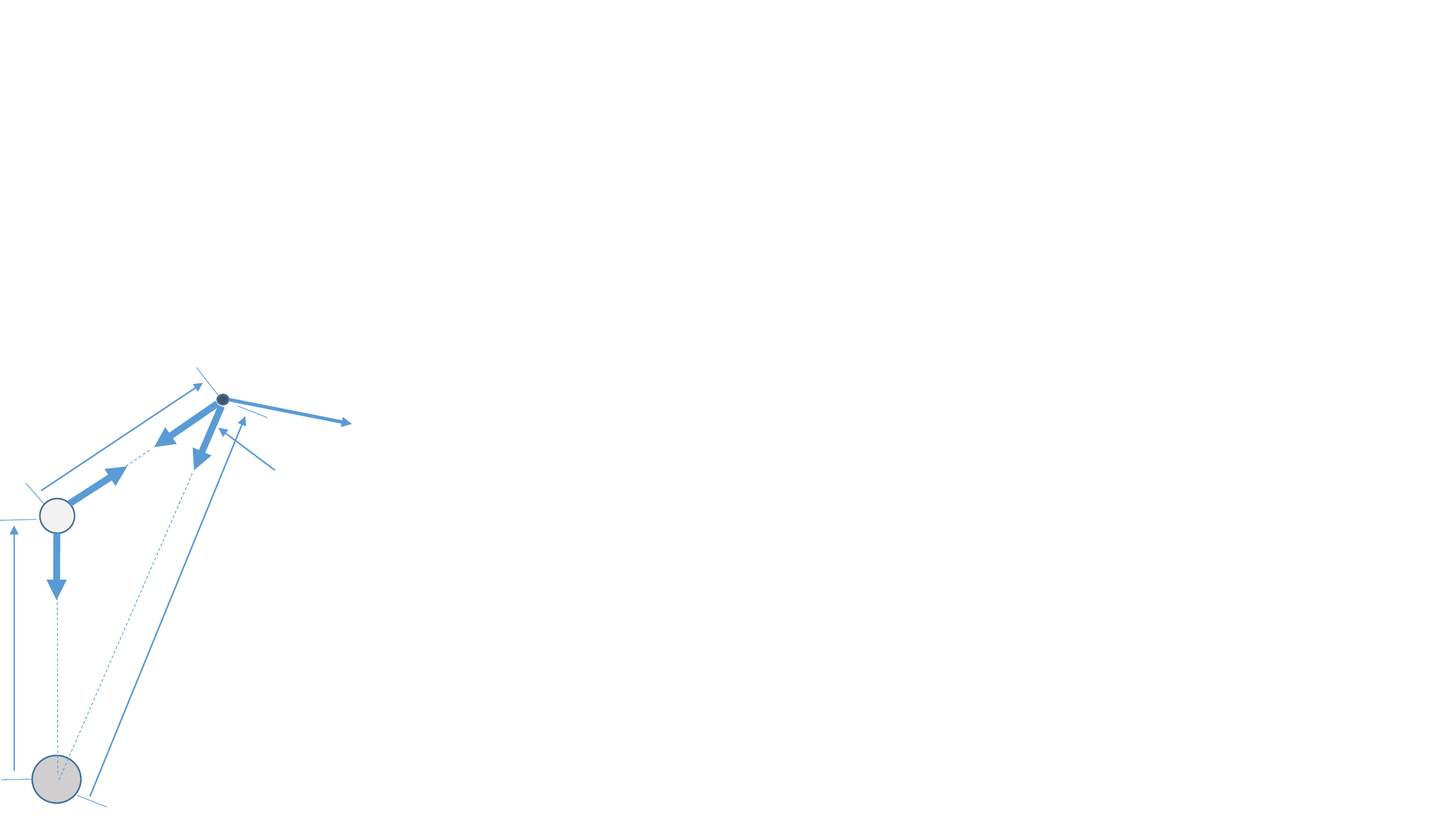}}
     \put(2.5,2.6){$\bR_c$}

    \put(.4,2.2){$\bR$}
     \put(1.5,5.6){$\br$}
     \put(-.5,.5){Sun}
     \put(-.8,4.5){asteroid}
     \put(3.5,6.3){spacecraft}
     \put(3.9,4.8){$\bF_{cs}$}
     \put(2.1,5.2){$\bF_{ca}$}
     \put(1.8,4.8){$\bF_{ac}$}
     \put(1,3.2){$\bF_{as}$}
     \put(4.6,5.5){$\bT$}
    \end{picture}
  \end{center}
  \caption{The asteroid-spacecraft system.}
\label{fbd}
\end{figure}
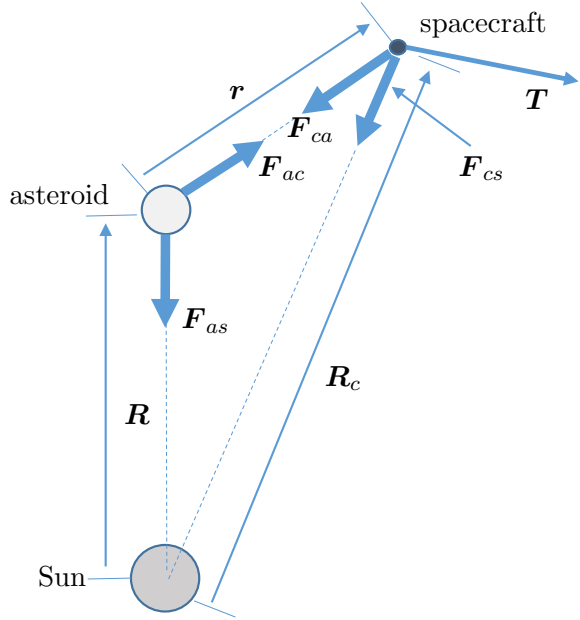

Consider a spacecraft flying near an asteroid in a known orbit about
the Sun as shown schematically in Figure \ref{fbd}. In the figure, the
position vector from the Sun to the asteroid is denoted by $\bR$,
the position vector from the asteroid to the spacecraft is denoted by $\br$,
and the position vector from the Sun to the spacecraft is denoted by
$\bR_c$. The gravitational
forces from the Sun on the asteroid and from the spacecraft on the
asteroid are denoted by $\bF_{as}$ and $\bF_{ac}$, respectively. The spacecraft is subjected to the gravitational forces
$\bF_{ca}$ and $\bF_{cs}$ from the asteroid and the Sun, respectively. In
addition, a propulsive force $\bT$ may act on the spacecraft from its
own thruster. Denoting the magnitudes of the vectors $\bR$, $\br$, and
$\bR_c$ by $R$, $r$ and $R_c$, respectively, the gravitational forces may be written as
\ben
\bF_{as}&=&-\frac{Gm_am_s}{R^3}\bR\\
\bF_{ac}&=&\frac{Gm_am_c}{r^3}\br
\label{fac}\\
\bF_{ca}&=&-\frac{Gm_cm_a}{r^3}\br\\
\bF_{cs}&=&-\frac{Gm_cm_s}{R_c^3}\bR_c\\
\een
where $m_s$, $m_a$, and $m_c$ are the masses of the Sun, the asteroid,
and the spacecraft, respectively, and $G$ is the universal
gravitational constant. It should be noted that the mass of the
spacecraft is time dependent, as it decreases with the burning of
fuel.

\subsection{The center of mass of the asteroid-spacecraft system}
In the following study of the effects of the forces on the asteroid-spacecraft
system in Figure \ref{fbd}, it will be useful to take into account the
location of the center of mass of the system. Thus the position of the center of mass of the
asteroid-spacecraft system, with respect to the center of the asteroid, may be expressed as
\be
\br_G=\frac{m_c}{m_a+m_c}\br,
\label{cm}
\ee
where $\br$ is defined in Figure \ref{fbd}. When the mass of the
spacecraft is much smaller than that of the asteroid ($m_c<<m_a$),
which will be assumed in what follows, the center of mass of the
system will be very close to the center of mass of the asteroid.
To get a sense for the typical distances involved, it is useful to
consider an example. Following \cite{lu-love} we
consider a hypothetical asteroid of diameter 100 m and mass 8.4 $\times$ 10$^9$
kg. Suppose that a
spacecraft of mass 2000 kg is at a distance of 100 m from the center
of the asteroid. It would follow from (\ref{cm}) that the center of
mass is at a distance of $r_G=2.4\times 10^{-5}$ m from the center of
the asteroid. In general, we conclude that when $m_c<<m_a$ the center
of mass of the system may be considered to be at the center of mass of
the asteroid. An analysis of the deflection of the asteroid may therefore
be based on either the motions of the center of mass of the asteroid itself, or equivalently, on the motions of the center of mass of the asteroid-spacecraft
system. In this paper the former approach will be taken.

\subsection{The thrust as an external force to the asteroid-spacecraft
  system}

It is suggested in Figure \ref{fbd} that the thrust $T$ only acts on the 
spacecraft and is therefore an external force to the asteroid-spacecraft system.
This would not be true if the generation of the thrust involved
an interaction with the asteroid (for example through significant forces between
the exhaust gases and the asteroid). It is therefore
important to validate this basic assumption that the thrust 
$\bT$ is indeed an external force to the system.

\begin{figure}[htbp]
  \begin{center}
    \unitlength=.5in
    \begin{picture}(4,4.7)
     \put(2.9,0){\includegraphics[scale=.8]{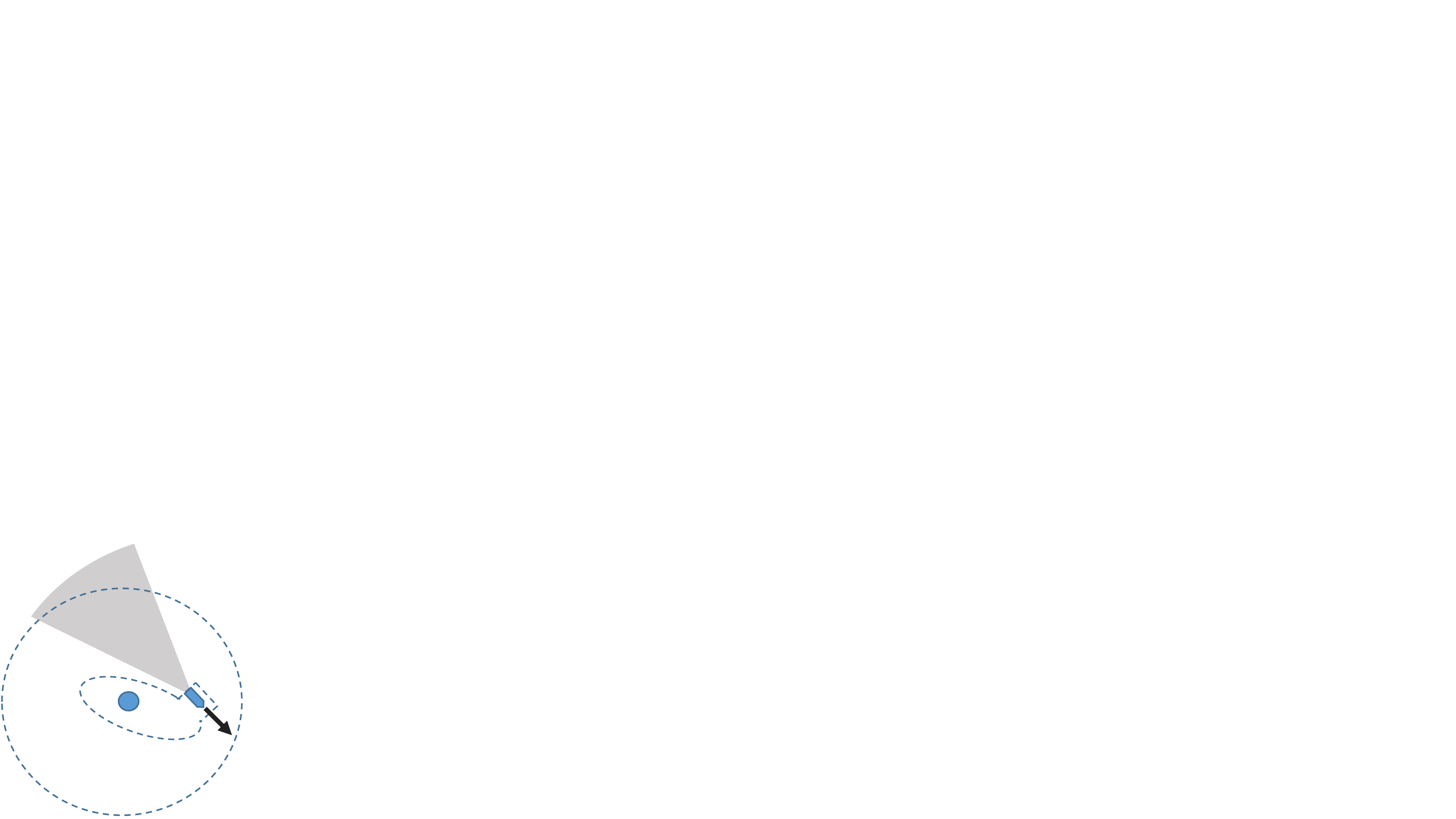}}
     \put(-2.7,0){\includegraphics[scale=.8]{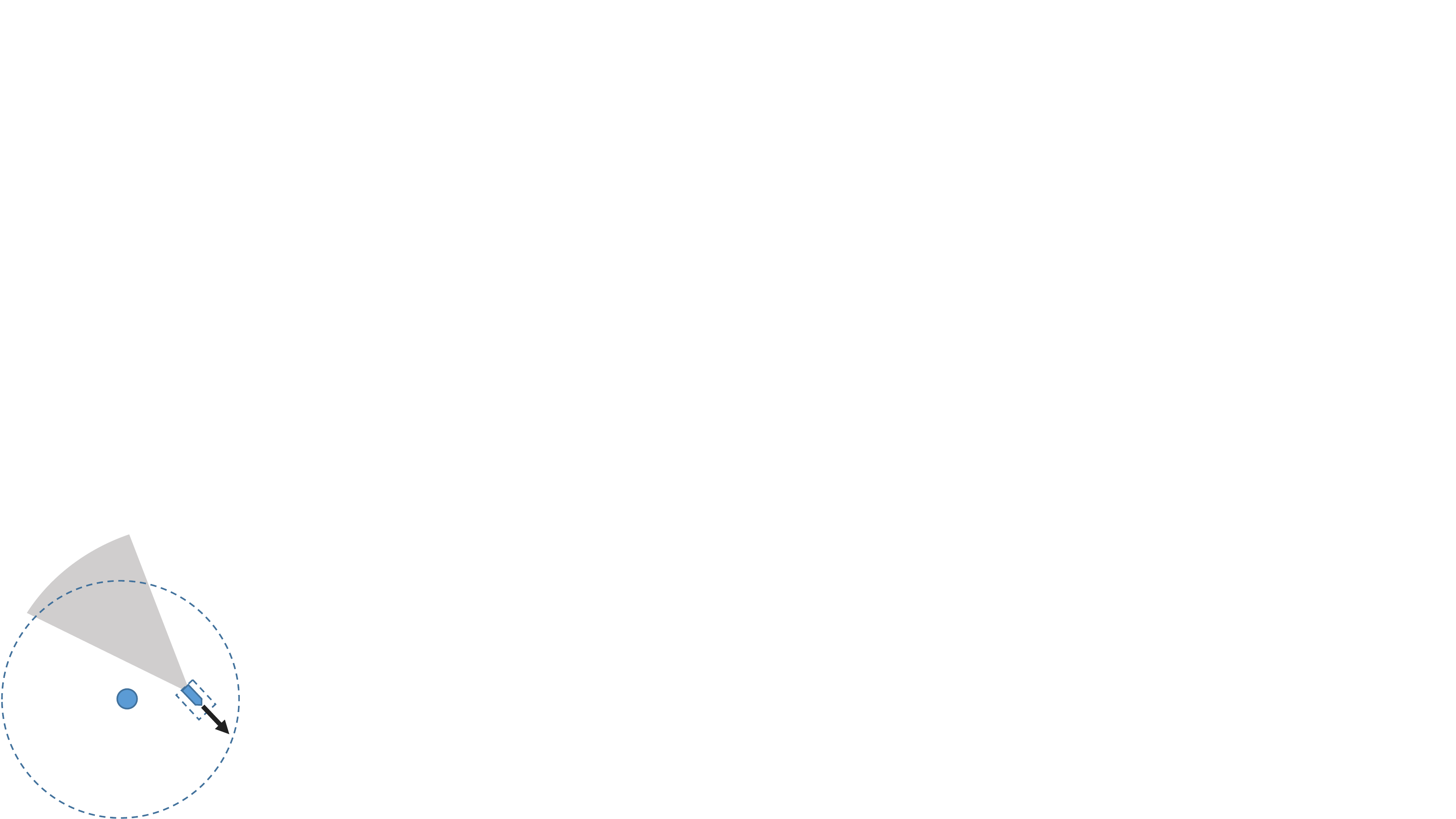}}
     \put(-2,1.9){asteroid}
     \put(-3.2,2.4){$SOI$}
     \put(5.9,1){$\bT$}
     \put(0.3,1.){$\bT$}
     \put(.3,2.0){$V_1$}
     \put(4,2.2){$V_2$}
     \put(-.7,4.1){Plume}
    \end{picture}
  \end{center}
  \caption{The asteroid-spacecraft system and control volumes.}
\label{momentum}
\end{figure}
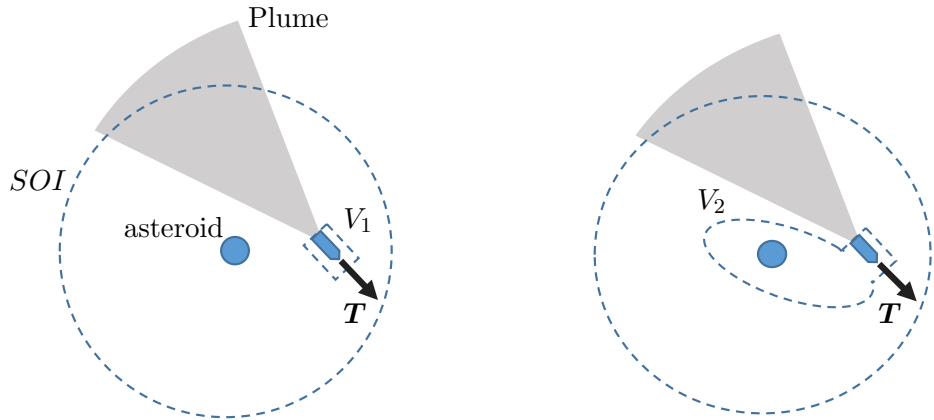

We first consider the thrust $\bT$ from the point of view of its
action on the spacecraft using a control volume 
approach \cite{greenwood} . Thus we define a volume $V_1$ that
encompasses the spacecraft as shown in Figure \ref{momentum}(a). The rate
of change of momentum inside the control volume can be related to the
net force on the mass inside the control volume (i.e. the total 
gravitational force on the spacecraft, say $\bF_c$) and
the rate at which mass flows in or out of the control volume. This
leads to the equation 
\be
\bF_c=\frac{d}{dt}\int_{V_1}\rho \bv dV+\int_{A_1}\rho \bv (\bv_r \cdot d\bA)
\label{int}
 \ee
where $\bF_c$ is the total external force on the spacecraft given by
\be
\bF_c=-\frac{Gm_am_c}{r^3}\br-\frac{Gm_sm_c}{R_c^3}\bR_c,
\ee
and $\int_{V_1}\rho \bv dV$ is the total momentum of all particles
within the control volume, i.e. the
spacecraft. The area $A_1$ is the surface area of the
control volume $V_1$ and $\bv_r$ is the component of the velocity of
the exhaust gases in a direction perpendicular to an area element
$d\bA$, with respect to the area element. The velocity $\bv_r$ is
defined positive if it corresponds to a flow of gases out of the
control volume.

The total momentum of the mass $m_c$ inside the
control volume can also be written in terms of the velocity of its 
center of mass $\bv_c$, i.e.
\be
m_c\bv_c=\int_{V_1}\rho \bv dV
\label{ma}
\ee
Rearranging the terms in (\ref{int}) and using (\ref{ma}) we have
\be
\frac{d}{dt}(m_c\bv_c)=\bF_c-\int_{A_1}\rho\bv (\bv_r\cdot d\bA)
\ee
or
\be
m_c\frac{d\bv_c}{dt}=\bF_c-\int_{A_1}\rho\bv (\bv_r\cdot d\bA)-\frac{dm_c}{dt}\bv_c
\ee
where the last two terms on the right hand side represent the thrust from
the rocket, i.e.
\be
\bT=-\int_{A_1}\rho\bv \bv_r\cdot d\bA-\frac{dm_c}{dt}\bv_c
\label{t1}
\ee

Once the exhaust gases are ejected from the spacecraft, they move at a
very large velocity with respect to the asteroid and out of the
$SOI$, as will be demonstrated by example below. Due to the low
density, and therefore small amount of total mass of gas within 
the $SOI$ at any time, the gravitational effect of the gases on the
asteroid can be neglected (just as their gravitational effects on the 
spacecraft may be neglected). With this in mind, we
next consider a larger control volume $V_2$, as in Figure
\ref{momentum}b, which consists of $V_1$ and an additional part that 
encompasses the asteroid. 
We can relate the total gravitational force on the
asteroid-spacecraft system to the rate of change of the momentum
inside the control volume in the form
\be
\bF=\frac{d}{dt}\int_{V_2}\rho \bv dV+\int_{A_2}\rho \bv (\bv_r \cdot d\bA)
\label{soi}
\ee
where $A_2$ is the total surface area of the control volume
$V_2$ and $\bF$ is the total external gravitational force on the asteroid-spacecraft system or
\be
\bF=-\frac{Gm_am_s}{R^3}\bR-\frac{Gm_cm_s}{R_c^3}\bR_c
\ee
Again the momentum within $V_2$ (i.e. the spacecraft and the
asteroid) may be written as
\be
\int_{V_2}\rho \bv dV=m_a\bv_a+m_c\bv_c
\label{ma2}
\ee
Using (\ref{soi}) and (\ref{ma2}), and noting that $m_a$ is constant we have
have
\be
m_a\frac{d\bv_a}{dt}+m_c\frac{d\bv_c}{dt}=\bF-\int_{A_2}\rho\bv (\bv_r\cdot d\bA)-\frac{dm_c}{dt}\bv_c
\ee
or denoting the velocity of the center of mass of the asteroid-spacecraft system by $\bv_G$
\be
(m_a+m_c)\frac{d\bv_{G}}{dt}=\bF-\int_{A_2}\rho\bv (\bv_r\cdot d\bA)-\frac{dm_c}{dt}\bv_c
\ee
where the last two terms on the right hand side constitute the thrust that is
generated on the asteroid-spacecraft system due to the ejection of
mass out of the volume $V_2$
\be
\bT_{syst}=-\int_{A_2}\rho\bv (\bv_r\cdot d\bA)-\frac{dm_c}{dt}\bv_c
\label{tsyst}
\ee
Noting that, from a physical point of view, the integrals in
(\ref{t1}) and (\ref{tsyst}) are identical leads to the 
conclusion that
\be
\bT=\bT_{syst}
\ee
and the thrust acting on the spacecraft can therefore be considered
an external force on the system that has the ability to perturb the
motion of the center of mass of the system about the Sun.

To further justify the assumption that the exhaust gases leave the
$SOI$ of the asteroid largely unaffected by the asteroid, it may be
useful to study an example. We
consider again an asteroid of diameter 100 m and mass 8.4 $\times$ 10$^9$
kg. The magnitude of the velocity of the exhaust with
respect to the spacecraft from a rocket of specific impulse $I_{sp}$
can be found from (see e.g. \cite{greenwood})
\be
v_{e/c}=gI_{sp}
\label{ve}
\ee
where $g=9.81$ m/s$^2$ is the acceleration due to gravity on the
surface of Earth. Thus for example with $I_{sp}=200$ s
\be
v_{e/c}=1962 {\mbox{m/s}}
\label{erc}
\ee

The velocity of a spacecraft in a circular orbit about the asteroid at
a radius of $r=75$ m would be
\be
v=\sqrt{\frac{\mu}{r}}=0.09 {\mbox{m/s}},
\label{circular}
\ee
therefore not significantly adding to the velocity of the exhaust.

On the other hand, the escape velocity near the surface of the asteroid considered is
\be
v_{esc}=\sqrt{\frac{2\mu}{r}}
\ee
or
\be
v_{esc}=0.15 {\mbox{m/s}}
\ee
It is therefore clear that, 
\be
v_e>>v_{esc}
\ee
due to the very large velocity of the
exhaust, it can always be expected to exit the $SOI$ of the asteroid
without further interaction with the asteroid.

Another force that in principle could be acting on the asteroid but that is
not included in the free body diagrams in Figure \ref{fbd}  is a
direct force from the exhaust plume from the spacecraft's thruster on the
asteroid. This force may exist if the exhaust plume impinges
on the asteroid, thus counteracting the effect of the gravitational force
from the spacecraft on the asteroid. In practice, this scenario can be easily avoided
by imposing a constraint on the possible directions of the thrust $T$, and only 
allowing those directions that do not lead to plume impingement, as is
suggested for example, in Figure \ref{momentum}.


\subsection{Equations of motion}
Considering the free body diagrams in Figure \ref{fbd}, the equation of motion for the asteroid and the spacecraft in the inertial frame may be written as
\ben
m_c\ddot{\bR_c}=\bF_{cs}+\bF_{ca}+\bT\\
\label{eom1}
m_a\ddot{\bR}=\bF_{as}+\bF_{ac}
\label{eom2}
\een
From (\ref{eom1}) it follows that the perturbing force on the asteroid
that will be responsible for a modification of its orbit is
$\bF_{ac}$, i.e. the gravitational force
from the spacecraft. In particular, the thrust $\bT$ does not
directly affect the motions of the asteroid. However, it serves the
important purpose of controlling the motions of the spacecraft with
respect to the asteroid, and therefore controlling the magnitude and
direction of $\bF_{ac}$. (See for example \cite{olympio} for the
alternative approach of considering the motions of the center of mass
of the system and where $\bT$ is the force that modifies the orbit of
the center of mass.)

\section{The asteroid deflection formula}
The equation of motion for the asteroid (\ref{eom1}) is in the general
form
\be
\frac{d\bp}{dt}=\bF_{as}+m_a\bA(t)
\label{strategy}
\ee
where
$\bp$ is the momentum of the asteroid and $\bA(t)$ is defined so that
\be
\bF_{ac}=m_a\bA(t)
\ee
or
\be
\bA=\frac{Gm_c}{r^3}\br
\label{ba}
\ee
The vector $\bA(t)$ may thus be interpreted as the perturbing
acceleration on the asteroid that is caused by the gravitational force
from the spacecraft.

\begin{figure}[htbp]
  \begin{center}
    \unitlength=.5in
    \begin{picture}(4,3)
     \put(-2,-1){\includegraphics[scale=0.8]{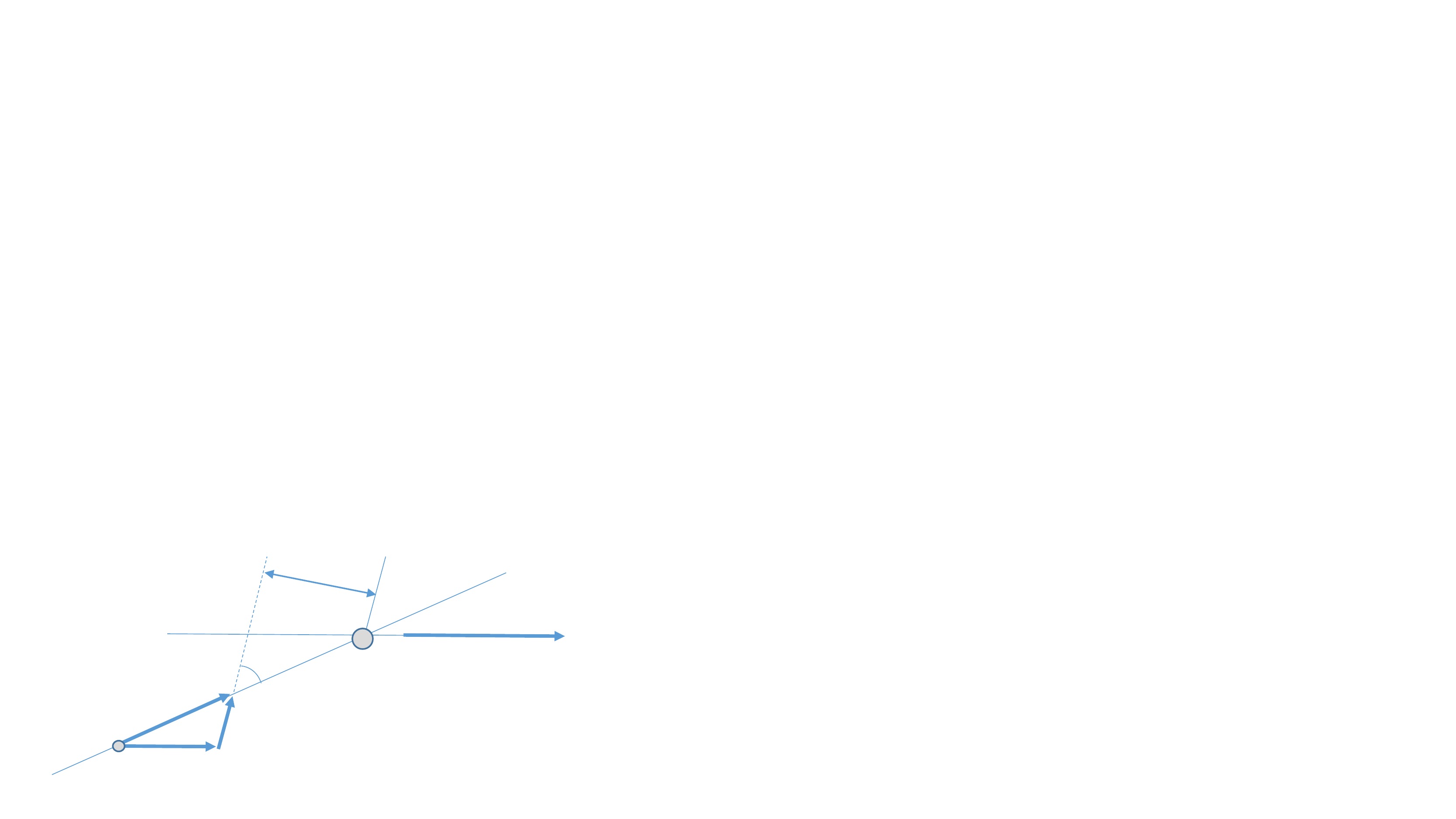}}
     \put(3,1.15){Earth}

    \put(5,1.85){$\bv_E$}
     \put(0,.6){$\bv_a(t_e)$}
     \put(1.4,.3){$\bv_{a/E}$}
     \put(-1.5,.1){asteroid}
     \put(2.5,2.1){$\Delta\zeta$}
     \put(1.7,1.3){$\psi$}
     \put(0.3,-.1){$\bv_E$}
    \end{picture}
  \end{center}
  \caption{Earth and the asteroid at the time of closest approach.}
\label{encounter}
\end{figure}
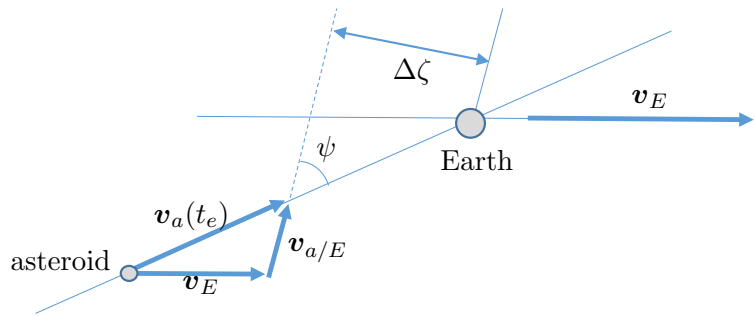

Denoting the time of projected encounter between Earth and the
asteroid by $t_e$, the goal for the acceleration $\bA$ is to effect
a non-zero distance of closest approach (i.e. a deflection)
between the Earth and the asteroid at that time.
This distance is described in Figure \ref{encounter} where it is denoted by $\Delta \zeta$ and measured in a direction perpendicular to the velocity of the asteroid with respect to Earth.

In general, the distance $\Delta\zeta$ at time $t_e$ will depend on the time $t_s<t_e$ when the 
perturbing acceleration $\bA$ begins to act and the time $t_f\leq t_e$ at
when stops acting. An expression for $\Delta \zeta$ known as the asteroid deflection formula is derived in \cite{izzo} and takes the form
\be
\Delta\zeta=\frac{3a}{\mu}v_a(t_e)\sin\psi\int_{t_s}^{t_f}(t_e-t)\bv_a\cdot\bA(t)dt
\label{deltazeta}
\ee
where $\bv_a$ is the time dependent heliocentric velocity of the
asteroid, $v_a(t_e)$ is the magnitude of the asteroid's velocity at the
time of the projected encounter, and $\psi$ is the angle between the
heliocentric velocity of the asteroid and its relative velocity with respect to
Earth (see Figure \ref{encounter}). Using (\ref{ba}) it follows that
\be
\Delta\zeta=\frac{3a}{\mu}v_a(t_e)\sin\psi\int_{t_s}^{t_f}(t_e-t)\bv_a\cdot\frac{Gm_c}{r^3}\br
dt
\label{deltazeta2}
\ee

Note that $\Delta \zeta$ can be positive or negative depending on the
time history of $\bv_a\cdot\bA$. The case $\Delta\zeta>0$ implies that the
asteroid arrives at the projected point of collision with Earth, after
the Earth has passed that point. This is the case depicted in Figure
\ref{encounter}. On the other hand, $\Delta \zeta<0$ would mean that
the asteroid passes the projected point of collision before the Earth
arrives at that point. Apart from the resulting sign of $\Delta \zeta$
there is no difference in how these two cases are analyzed. In what
follows we will assume that the required
deflection is
$\Delta \zeta<0$. Thus it follows from (\ref{deltazeta}) that
$\bA$ should be directed so that $\bv_a\cdot\bA<0$. In other words, the component of the 
gravitational force from the spacecraft on the asteroid should act so
as to oppose the heliocentric motion of the asteroid.

\begin{figure}[htbp]
  \begin{center}
    \unitlength=.5in
    \begin{picture}(4,3)
     \put(0,0){\includegraphics[scale=0.8]{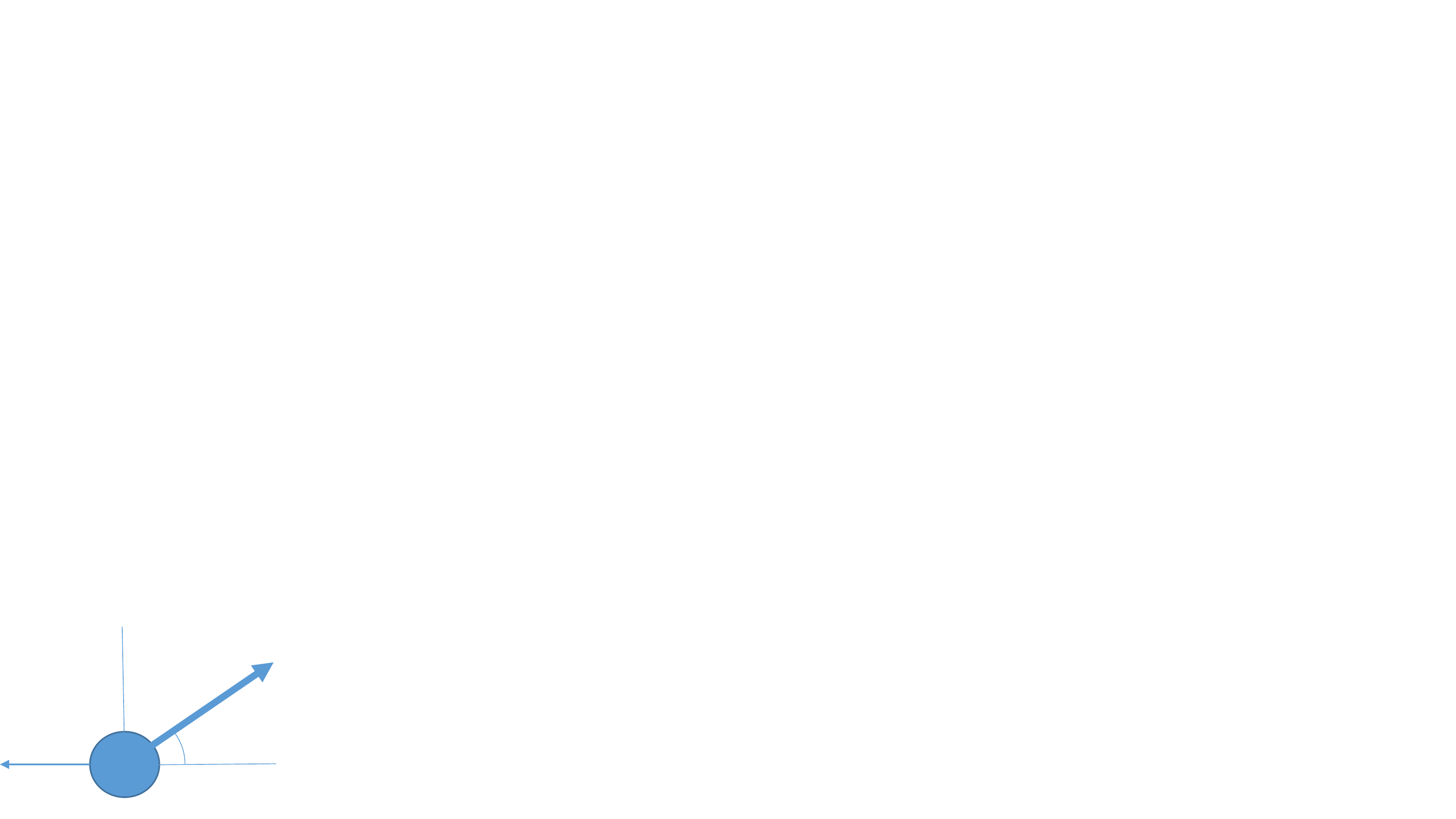}}
     \put(0,.5){$\bv_a$}
     \put(1,-.1){asteroid}
     \put(4.1,.8){$x$}
     \put(1.5,2.7){$y$}
     \put(2.8,1){$\theta$}
     \put(4,1.9){$\bF_{ac}$}
    \end{picture}
  \end{center}
  \caption{The gravitational force from the spacecraft on the asteroid
  in relation to the direction of the heliocentric velocity of the asteroid.}
\label{force}
\end{figure}
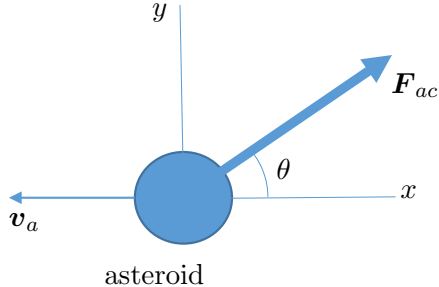

Figure \ref{force} illustrates the action of the gravitational force 
from the spacecraft on the asteroid in relation to the direction of
the asteroid's velocity, which is 
assumed to be pointing to the left in the figure. The $x$ axis is
defined to point in a direction opposite to the heliocentric velocity
of the asteroid $\bv_a$ and the $y$ axis points in a direction
perpendicular to $x$ and lies in the plane of the asteroid's orbit.
Denoting the $x$ component of the gravitational force
$\bF_{ac}$ by $F_{acx}$, and recalling (\ref{ba}) the condition that
$\bv_a\cdot\bA<0$ is 
equivalent to $F_{acx}>0$. In terms of the
angle $\theta$ between $\bF_{ac}$ and the $x$ axis, it is required that
 $|\theta|<\frac{\pi}{2}$.

  Of course, the direction of the gravitational force $\bF_{ac}$
is determined by the position of the spacecraft with respect to the
asteroid as this force always points toward the spacecraft.  This implies that the spacecraft
should always be in the right half of the $xy$ plane as defined in
Figure \ref{force}. In the next section, it will be described how the spacecraft can maintain this
type of motion.




\subsection{Restricted Keplerian Motion}
\begin{figure}[htbp]
  \begin{center}
    \unitlength=.5in
    \begin{picture}(4,3)
     \put(-5.9,-4.5){\includegraphics[scale=0.5]{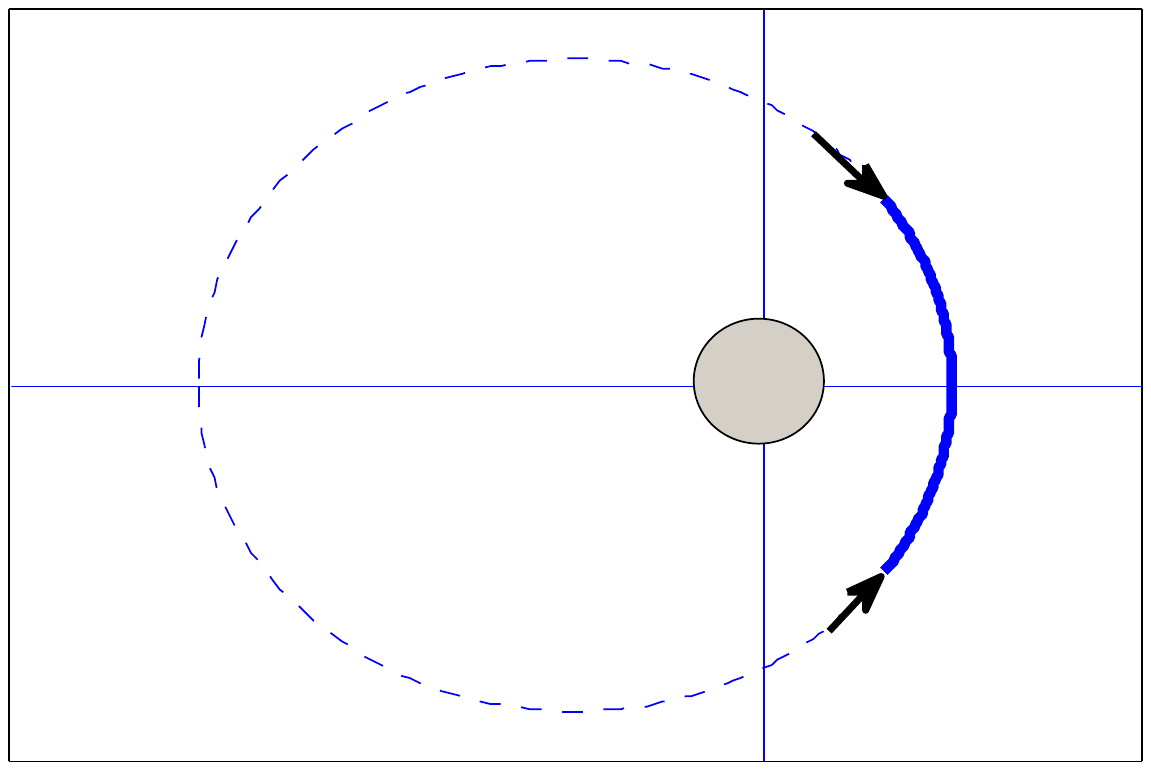}}
     \put(.9,-4.5){\includegraphics[scale=0.5]{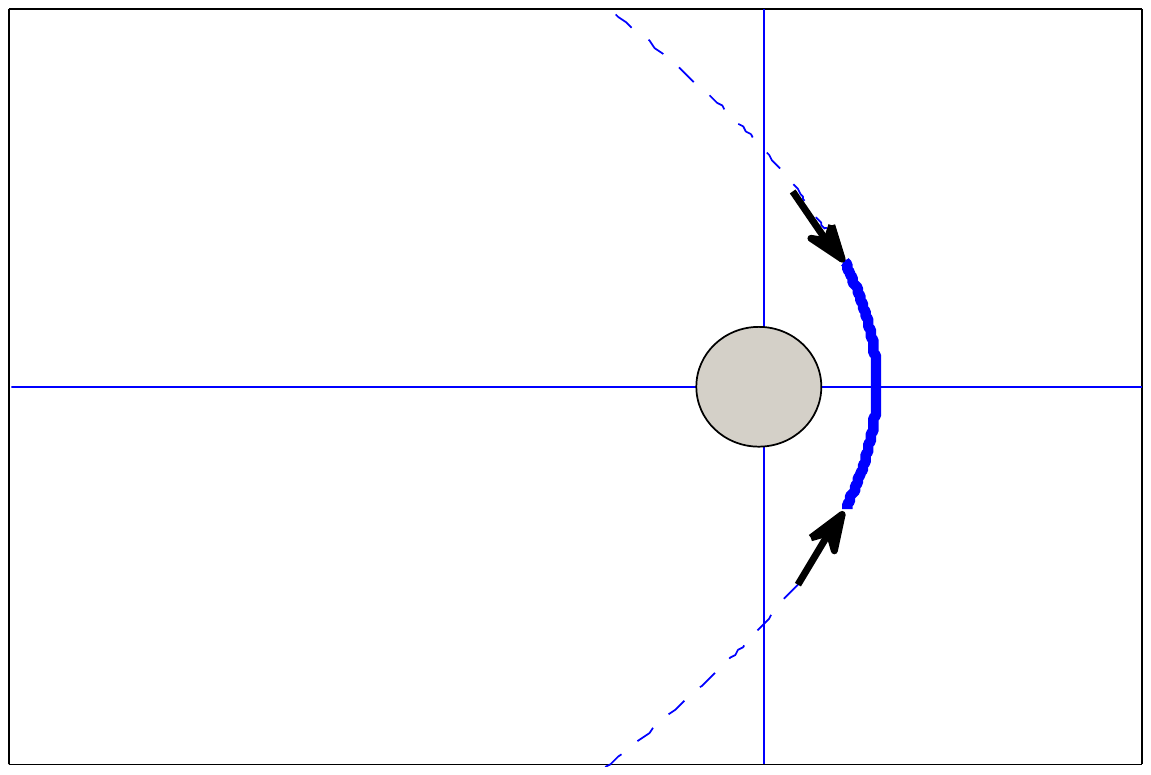}}
     \put(-.4,2){$\Delta v$}
     \put(-.5,-.1){$\Delta v$}
     \put(6.2,.1){$\Delta v$}
     \put(6.2,1.8){$\Delta v$}
     \put(-2.2,.8){asteroid}
     \put(-3.6,-.1){(a)}
     \put(3.2,-.1){(b)}
    \end{picture}
  \end{center}
  \caption{Schematic illustration of restricted Keplerian motion: (a)
    on an elliptic orbit, (b) on a hyperbolic orbit.}
\label{segments}
\end{figure}

When the spacecraft is within the sphere of influence (SOI) of the
asteroid, its motion with respect to the asteroid may be studied using
the two-body model. It is then well known that any unforced motion
will take place along a conic section (ellipse,
parabola, or hyperbola) that will generally visit both halves of the
plane. Thus, in order to keep the spacecraft in the right half plane, so that
$|\theta|<\pi/2$ 
in Figure \ref{force}, it will be  necessary to restrict its motion by 
imparting intermittent impulsive velocity changes $\Delta v$. In this way,
the spacecraft can be made to move along a segment of a conic section
in alternating directions. Figure
\ref{segments}  shows two examples of such motion. In Figure \ref{segments}(a)
the spacecraft moves back and forth along an elliptic orbit, and
in  Figure \ref{segments}(b) the spacecraft moves back and forth along a
hyperbolic orbit. A similar motion is of course possible along a
parabolic orbit as well. The changes in velocity $\Delta v$ qualitatively depicted in Figure
\ref{segments} are imparted through the action of the thrust $\bT$ on
the spacecraft. In what follows, the motion of the back and forth
motion of the spacecraft as described above will be referred to as
``restricted Keplerian motion'' and the corresponding gravity tractor
will be referred to as a  ``Keplerian'' gravity tractor. 

It is worth noting that the $\Delta v$'s required for such motions
around large objects such as planets would be prohibitively large due
to the generally large velocities of the spacecraft that would be
needed. The orbital
velocities around an asteroid, on the other hand, are generally small as
has been exemplified in (\ref{circular}) making this restricted
Keplerian motion plausible.

\subsection{An expression for the resulting deflection: the average force}
While Figure \ref{segments} describes qualitatively the motion of the
spacecraft with respect to the asteroid, the orbit itself, as defined by its
parameters such as semimajor axis and eccentricity, and the length of the
segment on the orbit along which the spacecraft will travel, will
influence the overall deflecting effect of the gravitational force
from the spacecraft. The dependence of the performance of the gravity
tractor on these parameters will be studied in this and the following sections.

\begin{figure}[htbp]
  \begin{center}
    \unitlength=.5in
    \begin{picture}(4,5)
     \put(2.4,2.3){$\theta$}
     \put(-1.6,-1){\includegraphics[scale=0.55]{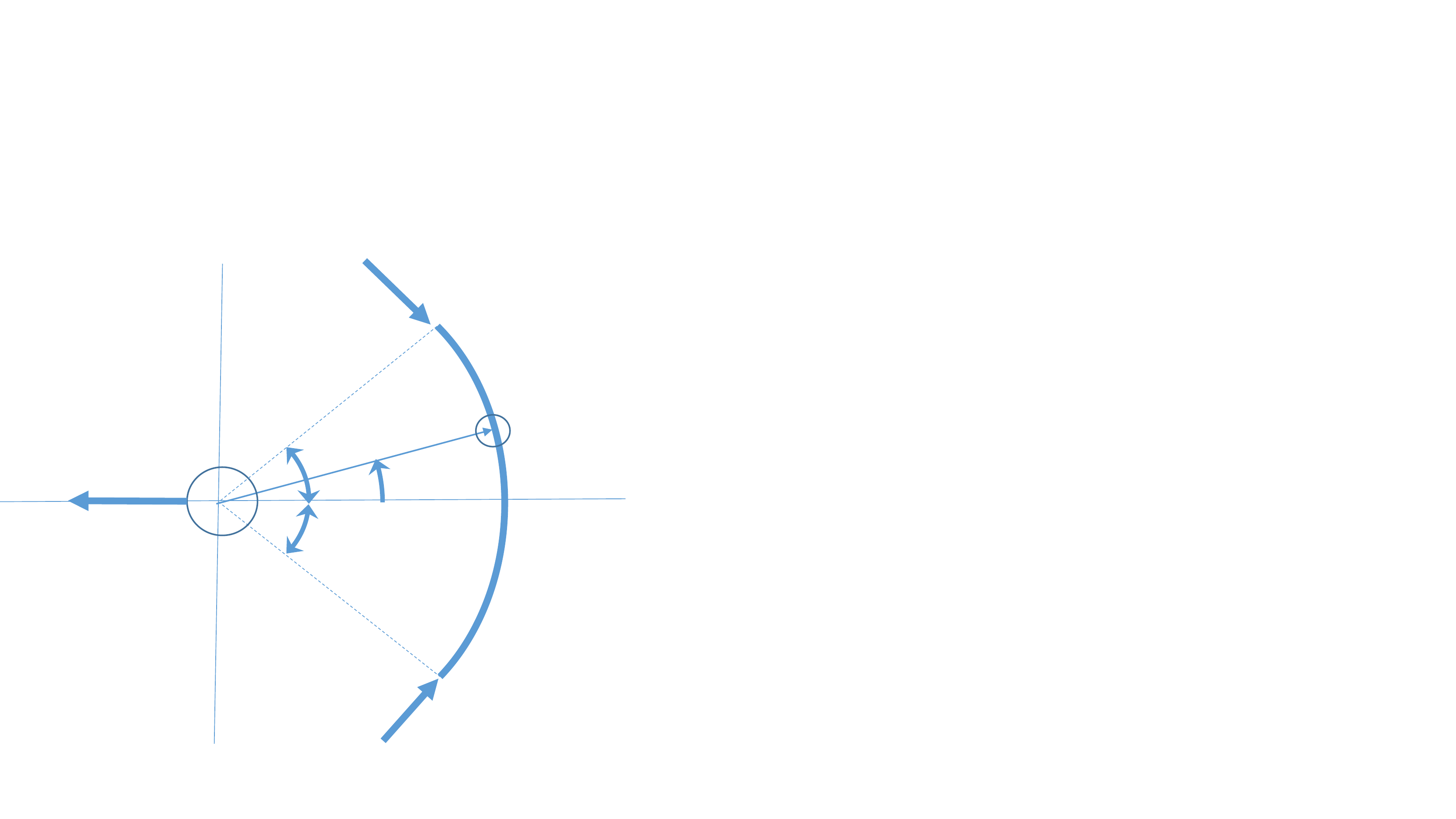}}
    \put(1.5,2.6){$\theta_b$}
     \put(1.5,1.7){$\theta_b$}
     \put(2.7,2.8){$r$}
     \put(-.5,2.6){asteroid}
     \put(-.8,1.8){$v_a$}
     \put(3.6,3.){spacecraft}
     \put(2.1,4.7){$\Delta v_2$}
     \put(1.9,.1){$\Delta v_1$}
     \put(3,.2){$A$}
     \put(2.9,4.1){$B$}
     \put(4.9,2.1){$x$}
     \put(.4,4.7){$y$}

    \end{picture}
  \end{center}
  \caption{The asteroid-spacecraft system.}
\label{system}
\end{figure}
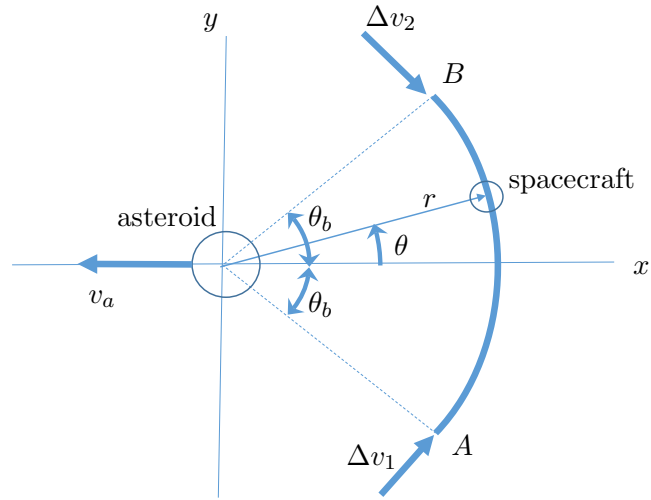

Figure \ref{system} depicts schematically a segment of a generic conic section along which
the spacecraft flies back and forth in a restricted Keplerian
motion. Thus the spacecraft flies from point $A$ to point $B$, then
back to point $A$ etc. This motion is made possible through impulsive
thrusts $\Delta v_1$ and $\Delta v_2$ that change the direction of the
velocity of the spacecraft at points $A$ and $B$, respectively. The
maximum $|\theta|$ at which the impulsive thrusts are delivered will
be denoted by $\theta_b$ and will be referred to as the bounding
angle in what follows.

The time it takes the spacecraft to go from point $A$ to point $B$, or
subsequently from point $B$ to point $A$, can be calculated using the the time
of flight formulas for the two body problem (see Section 4). Here we
just note that the time is independent of the direction of motion and
only
depends on the bounding angle $\theta_b$ and parameters of the
given conic section, namely the semimajor axis and eccentricity. 

The total deflection caused by this back and forth motion can be
calculated using the deflection formula (\ref{deltazeta2}). With the
help of Figure \ref{system} we note that 
\be
\bv_a\cdot\frac{Gm_c}{r^3}\br=-v_a\frac{Gm_c}{r^2}\cos\theta
\ee
where $v_a$ is the  (generally time dependent) magnitude of
$\bv_a$. Using this expression in (\ref{deltazeta2}) leads to
\be
\Delta\zeta=-\kappa\int_{t_s}^{t_f} (t_e-t)v_a\frac{Gm_c}{r^2}\cos\theta dt\;,
\label{dzetatot}
\ee
where we have defined for brevity
\be
\kappa=\frac{3a}{\mu}v_a(t_e)\sin\psi
\label{kappa}
\ee

To evaluate the integral in (\ref{impab1}) we note that the interval
of integration $[t_s,t_f]$ may be broken up into smaller intervals, each
corresponding to one pass of the spacecraft along the orbit segment in
Figure \ref{system}, either going from $A$ to $B$ or from $B$ to
$A$. Suppose that there are $N$ such passes on the orbit segment
during the time interval $[t_s,t_f]$. The
successive times at which the spacecraft reaches the
endpoints of the orbit segment may then be denoted by 
$t_i: i=1,2, \dots t_{N+1}$ 
where $t_1=t_s$ (the initial time), $t_{N+1}=t_f$, and
\be
t_{i+1}=t_i+\Delta t
\ee
where $\Delta t$ is the time of flight between points $A$ and $B$.
It then follows that the integral in (\ref{dzetatot}) may be written as
\be
\Delta\zeta=\sum_{i=1}^N\Delta\zeta_i
\label{totaldeltaz}
\ee
where $\Delta\zeta_i$ is the contribution to $\Delta\zeta$ from the
time interval $[t_i,t_{i+1}]$ that is given by
\be 
\Delta\zeta_i=-\kappa\int_{t_i}^{t_{i+1}} (t_e-t)v_a\frac{Gm_c}{r^2}\cos\theta dt\;,
\label{impab1}
\ee
Note that the mass of the spacecraft $m_c$ is constant within the
time interval $[t_i,t_{i+1}]$ as the $\Delta v$'s are executed at the
  times $t_i$. This constant mass in the time interval $[t_i,t_{i+1}]$ will be
  denoted by $m_{ci}$. Also, the integrand in (\ref{impab1}) can be simplified by using the definition of the magnitude of the angular momentum for the spacecraft about the asteroid (see e.g. \cite{curtis2,prussing1})
\be
h=r^2|\dot\theta|,
\label{ech}
\ee
Using (\ref{ech}) and the constant mass of the spacecraft $m_{ci}$ in
(\ref{impab1}) leads to 
\be
\Delta\zeta_i=-\kappa \int_{t_i}^{t_{i+1}} (t_e-t)v_a\frac{Gm_{ci}}{h }\cos\theta |\dot{\theta}| dt\;,
\label{impab2}
\ee
Suppose now that in a time interval $[t_i,t_{i+1}]$ the spacecraft is moving along the orbit segment in
the counterclockwise direction, i.e. $\dot\theta>0$. Then the true anomaly $\theta$ will vary from
$\theta=-\theta_b$ to $\theta=\theta_b$. In this case, (\ref{impab2})
may be written
\be
\Delta\zeta_i=-\kappa \int_{t_i}^{t_{i+1}} (t_e-t)v_a\frac{Gm_{ci}}{h }\cos\theta \dot{\theta} dt\;,
\label{impab3}
\ee
To further simplify (\ref{impab3}), we consider the order of magnitude of the time interval $[t_i,t_{i+1}]$. Thus revisiting the asteroid-spacecraft
system described at the end of Section 2.2, we note that the time it
would take the spacecraft to fly through an angle of $\pi$ radians,
for example, along a circular
orbit of radius of r=100 m would be (see e.g. \cite{curtis3})
\be
\Delta t=\pi \sqrt{\frac{r^3}{\mu}}
\ee
or $\Delta t=4200$ s or about 70 minutes. This time may be used as a characteristic value for a time interval $[t_i,t_{i+1}]$.
On the other hand, the velocity of the asteroid varies slowly
over much larger time intervals -- in the order of one year.
Thus, considering that $\Delta t/(1 \mbox{year})$ is in the order of $1\times10^{-4}\times$, the velocity $v_a$ may be taken to have a constant
value $v_{ai}$ in a time interval $[t_i,t_{i+1}]$. This allows for the
direct evaluation of the integral in (\ref{impab3}). Integrating by parts, and noting that $\theta=-\theta_b$ at time $t=t_i$, and that  $\theta=\theta_b$ at time $t_{i+1}$,  this leads to 
\be
\Delta \zeta_i =-\kappa \frac{Gm_{ci}}{h}v_{ai}\left[(t_e-t_{i+1}))\sin\theta_b+(t_e-t_{i}))\sin\theta_b\right]+\kappa \frac{Gm_{ci}}{h}v_{ai}\int_{t_i}^{t_{i+1}}\sin\theta dt
\label{dzi0}
\ee
We note that
\be
\int_{t_i}^{t_{i+1}}\sin\theta dt=0
\ee
because $\sin\theta$ is antisymmetric about the midpoint of the
interval $[t_i,t_{i+1}]$. Next, simplifying (\ref{dzi0}) it follows
that
\be
\Delta \zeta_i =-\kappa v_{ai}\frac{Gm_{ci}}{h}(2t_e-(t_{i+1}+t_i))\sin\theta_b
\label{dzi2}
\ee

For the case $\dot{\theta}<0$, the above procedure is repeated with
the only difference that $|\dot{\theta}|=-\dot{\theta}$ and that
$\theta=\theta_b$ at time $t=t_i$, and that  $\theta=-\theta_b$ at
time $t_{i+1}$. The result is again (\ref{dzi2}), i.e. the deflection
caused by the motion in the time interval $[t_i,t_{i+1}]$ is independent
of whether the spacecraft is moving clockwise or counter-clockwise.

A simpler form for $\Delta \zeta_i$ can be obtained by defining the time 
\be
\bar{t}_i=\frac{t_{i+1}+t_i}{2}
\label{bart}
\ee
and noting that
\be
\Delta \zeta_i =-\kappa v_{ai}(t_e-\bar{t}_i)\frac{2\sin\theta_bGm_{ci}}{h}
\label{dz3}
\ee

\subsubsection{The impulse exerted and the average force}
For a simplifying interpretation of (\ref{dz3}) we next consider
the impulse imparted to the asteroid by the gravitational force
from the spacecraft during one pass of the spacecraft along the
segment of the Keplerian orbit, i.e. during the time interval
$[t_i,t_{i+1}]$. Note that due to the symmetry of the orbit segment
about the $x-$axis in Figure \ref{system}, the impulse will only have
an $x-$component, and may be written as
\be
I_{i}=\int_{t_i}^{t_{i+1}} \frac{Gm_am_{ci}}{r^2}\cos\theta dt\;
\label{impulse}
\ee
Using (\ref{ech}), (\ref{impulse}) can be written as
\be
I_{i}=\int_{t_i}^{t_{i+1}} \frac{Gm_am_{ci}}{h}\cos\theta |\dot{\theta}|dt\;
\label{ii1}
\ee
or
\be
I_{i}=\frac{2Gm_am_{ci}}{h}\sin\theta_{b}\;.
\label{ii2}
\ee
Also, for later use, we note that the angular momentum can be written 
in terms of the periapsis
distance $r_p$ and the eccentricity $e$ of the orbit, i.e. (cf. \cite{curtis2,prussing1})
\be
h=\sqrt{\mu_ar_p(1+e)}\;,
\ee
where $\mu_a=Gm_a$ is the gravitational parameter for the asteroid. This
leads to the expression
\be
I_{i}=\frac{2Gm_am_c}{\sqrt{\mu_ar_p(1+e)}}\sin\theta_b\;.
\label{impab4}
\ee
Next, using (\ref{ii2}) it follows that (\ref{dz3}) can be written in the form 
\be
\Delta \zeta_i =-\frac{\kappa}{m_a} (t_e-\bar{t}_i)I_{i}v_{ai}
\label{dzinew}
\ee
and the total $\Delta\zeta$ is then obtained as
\be
\Delta \zeta =-\frac{\kappa}{m_a} \sum_{i=1}^N(t_e-\bar{t}_i)I_{i}v_{ai}
\label{deltazetaimp}
\ee
Defining the average force that acts during the time interval $[t_i,t_{i+1}]$ by
\be
\bar{F}_{i}=\frac{I_i}{\Delta t}
\label{barfi}
\ee
where 
\be
\Delta t=t_{i+1}-t_i
\label{deltat}
\ee
we may write $\Delta\zeta_i$ in the form
\be
\Delta \zeta_i =-\frac{\kappa}{m_a} (t_e-\bar{t}_i)\bar{F}_{i}v_{ai}\Delta t
\label{dzi}
\ee
Now, using (\ref{dzi}) in (\ref{totaldeltaz}) the total deflection can be
written as
\be
\Delta \zeta=-\frac{\kappa}{m_a} \sum_{i=1}^{N}(t_e-\bar{t}_i)\bar{F}_{i} v_{ai}\Delta t 
\label{deltazetafinal}
\ee

\subsection{Comparison to deflection from a constant force: the stationary and displaced-orbit gravity tractors}
The deflection that can be realized using a Keplerian gravity tractor may be compared to the corresponding deflections that would result from a stationary \cite{lu-love,olympio} or displaced-orbit \cite{mcinnes2} gravity tractors.

\subsubsection{The stationary gravity tractor}
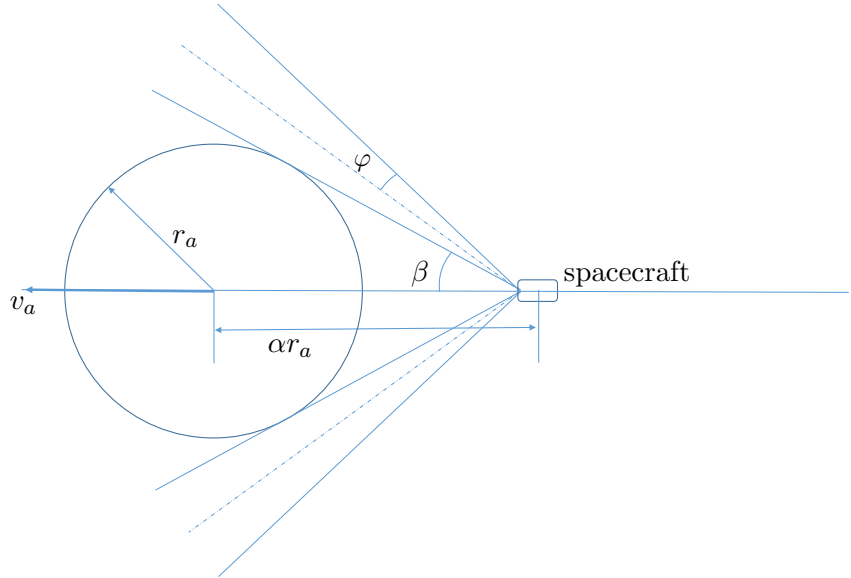
\begin{figure}[htbp]
  \begin{center}
    \unitlength=.5in
    \begin{picture}(4,5.5)
     \put(2.2,2.6){$\beta$}
     \put(-2,-.5){\includegraphics[scale=0.4]{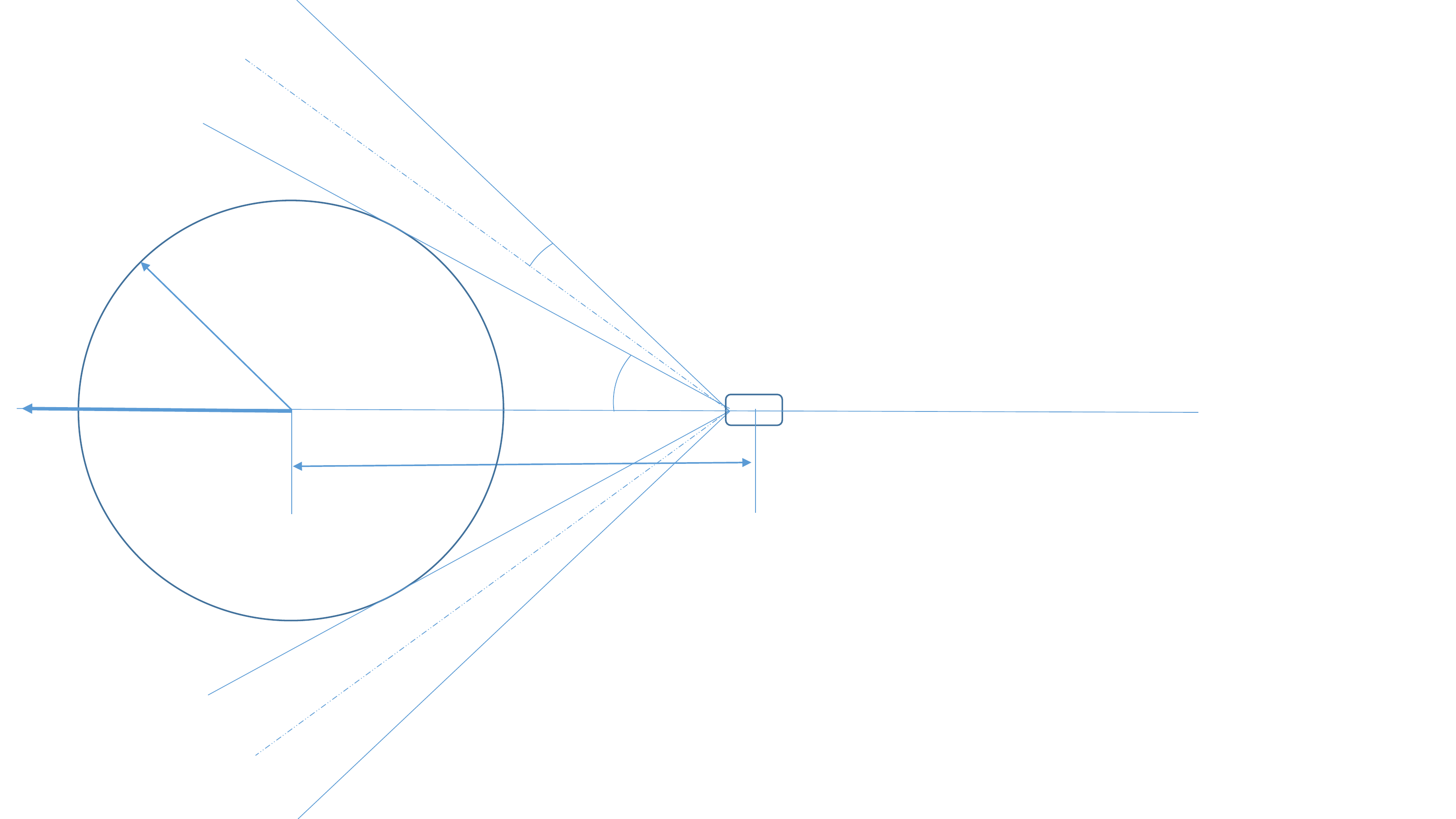}}
    \put(-.3,3){$r_a$}
     \put(-2,2.3){$v_a$}
     \put(3.8,2.6){spacecraft}
     \put(.7,1.85){$\alpha r_a$}
     \put(1.6,3.8){\small{$\varphi$}}

    \end{picture}
  \end{center}
  \caption{The stationary gravity tractor.}
\label{fig:stat}
\end{figure}

For the stationary
gravity tractor, referring to Figures \ref{fig:stat}, the corresponding force
exerted on the asteroid by a spacecraft of  mass ($m_c$) is
\be
F_s=\frac{G m_{c}m_a}{(\alpha r_a)^2}
\label{etats}
\ee
where $\alpha>1$ is a parameter that determines the distance to the
spacecraft from the center of mass of the asteroid, and $r_a$ is the
radius of the asteroid, as before. 

In this
case, it follows from (\ref{impab1}) that the contribution to the
deflection $\Delta\zeta$ at the time of 
projected collision $t_e$ from the constant gravitational force acting in the time interval $[t_i,t_{i+1}]$ can be calculated as
\be
\Delta\zeta^c_i=-\kappa \int_{t_i}^{t_{i+1}} (t_e-t)v_{ai}\frac{Gm_c}{(\alpha r_a)^2}dt\;,
\label{constf}
\ee
While the mass of the spacecraft will decrease slowly over time as it
burns fuel to generate the required thrust, this change in mass may be
neglected in the small time interval
$[t_i,t_{i+1}]$. Thus denoting the mass of the spacecraft in that time
interval by $m_{ci}$ we have from (\ref{constf})
that
\be
\Delta\zeta^c_i=\frac{\kappa}{2}\left[(t_e-t_{i+1})^2
  -(t_e-t_i)^2\right]v_{ai}\frac{Gm_{ci}}{(\alpha r_a)^2}
\label{eval}
\ee
Expanding and simplifying the term within square brackets of the right
hand side of (\ref{eval}) gives
\be
\Delta\zeta^c_i=\frac{\kappa}{2}\left[2t_e(t_i-t_{i+1})+t_{i+1}^2-t_i^2\right]v_{ai}\frac{Gm_{ci}}{(\alpha r_a)^2}
\label{eval1}
\ee
or
\be
\Delta\zeta^c_i=\frac{\kappa}{2}(t_{i+1}-t_i)\left[t_{i+1}+t_i-2t_e\right]v_{ai}\frac{Gm_{ci}}{(\alpha r_a)^2}
\label{eval2}
\ee
Now using (\ref{bart}), (\ref{deltat}), and (\ref{fac}), (\ref{eval2}) may be written as
\be
\Delta\zeta^c_i=-\frac{\kappa}{m_a}(t_e-\bar{t}_i)\Delta tv_{ai}F^i_{s}
\label{eval3}
\ee
where the force from the stationary gravity tractor in the $i$th time interval is 
\be
F^i_{s}=\frac{Gm_am_{ci}}{(\alpha r_a)^2}
\label{fis}
\ee
is the force on the asteroid from the stationary gravity tractor
during the time interval $[t_i,t_{i+1}]$.

\subsubsection{The displaced-orbit gravity tractor}
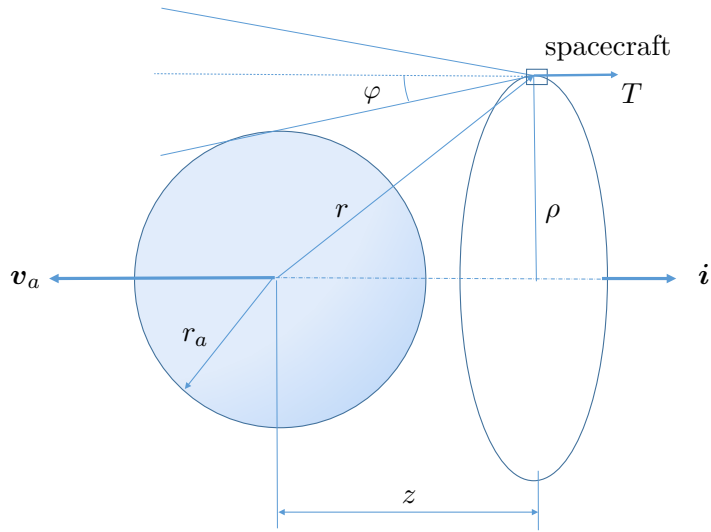
\begin{figure}[htbp]
  \begin{center}
    \unitlength=.5in
    \begin{picture}(4,5.5)
     \put(2.2,2.6){$\beta$}
     \put(-2,-.5){\includegraphics[scale=0.4]{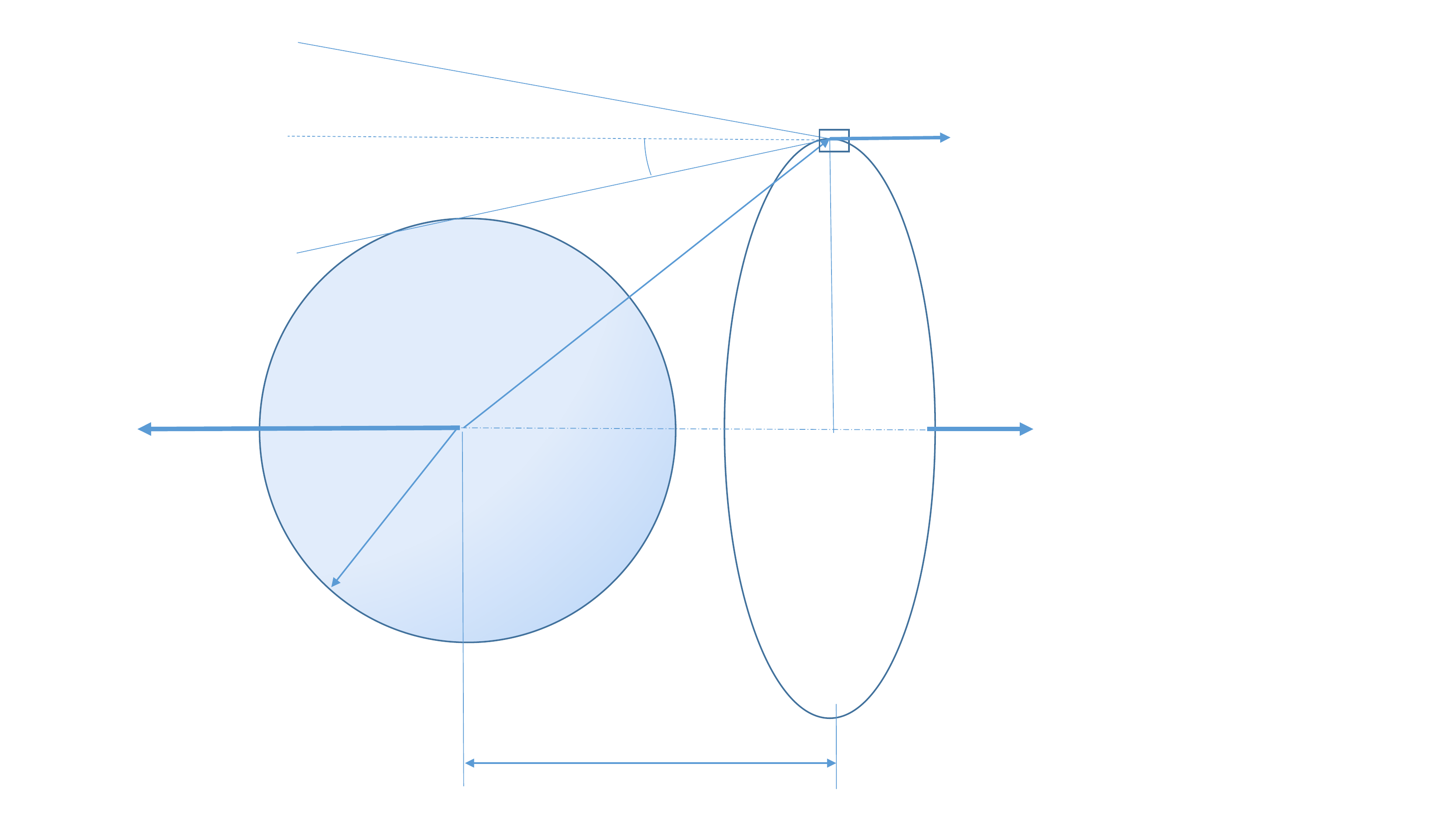}}
    \put(0.4,1.7){$r_a$}
    \put(2,3){$r$}
     \put(-1.4,2.3){$\bv_a$}
     \put(5.8,2.3){$\bi$}
     \put(4.2,4.7){spacecraft}
     \put(5,4.2){$T$}
     \put(4.2,3){$\rho$}
     \put(2.7,0){$z$}
     \put(2.3,4.25){\small{$\varphi$}}
    \end{picture}
  \end{center}
  \caption{The displaced-orbit gravity tractor.}
\label{fig:dogt}
\end{figure}

The displaced-orbit gravity tractor is shown schematically in Figure \ref{fig:dogt}.
The net perturbing gravitational force that acts on the asteroid from the spacecraft is  (cf. \cite{mcinnes2})
\be
\bF_d=\frac{Gm_am_c}{r^2}\frac{z}{r}{\bi}
\label{eff} 
\ee
and acts in a direction opposite to the velocity.
The corresponding required thrust on the spacecraft that keeps it in its orbital plane with respect to the asteroid is 
\be
T=  F_d=\frac{Gm_am_c}{r^2}\frac{z}{r} 
\label{tee}
\ee
It further follows from Figure \ref{fig:dogt} that
\be
r=\sqrt{\rho^2+z^2}
\label{r}
\ee
and that
\be
\rho=r_a+z\tan\beta
\ee
or 
\be
r=\sqrt{(r_a+z\tan\varphi)^2+z^2}
\label{arr}
\ee
It is described \cite{mcinnes2} that the largest perturbing gravitational force that can be exerted on the asteroid by the spacecraft while satisfying the 
non-impingement conditions for $\varphi=20^{o}$ occurs for $z=2.1 r_a$.
Using (\ref{arr}) and (\ref{eff}) this implies 
\be
F_{d}=0.21 \frac{Gm_am_c}{r_a^2}\;.
\label{fd}
\ee
The force $F_d$ will slowly decrease as $m_c$ decreases. However, in the small time interval $[t_i,t_{i+1}]$ it may be assumed constant, corresponding to the mass of the spacecraft in that time interval. Thus for the force in the $i$th time interval we have
\be
F^i_{d}=0.21 \frac{Gm_am_{ci}}{r_a^2}\;.
\label{fdi}
\ee
As in the case of the stationary gravity tractor, the contribution to the deflection at time $t_e$ from the action of $F^i_d$ in the time interval $[t_i,t_{i+1}]$ may be written as
\be
\Delta\zeta^d_i=-\frac{\kappa}{m_a}(t_e-\bar{t}_i)\Delta tv_{ai}F^i_{d}
\label{eval4}
\ee

The contributions to the deflection from a Keplerian gravity tractor,
the stationary gravity tractor, and the displaced orbit gravity
tractor from action during a time interval $[t_i,t_{i+1}]$ have
identical expressions save for the magnitudes of the forces (average force
in the case of the Keplerian gravity tractor) for a given mass of the
spacecraft $m_{ci}$. A measure of their relative efficiency may
therefore be obtained by a direct comparison of the forces. This
comparison will be made in Section 4.

\subsection{An expression for the total deflection}
Finally in this section we derive an expression for the total
deflection of the asteroid in terms of the mass of the spacecraft and
the amount of fuel burned. This expression will be obtained by
directly evaluating (\ref{deltazetafinal}).

Using the definition of $\Delta t$ in (\ref{deltat}) we note that
\ben
t_i=t_s+(i-1)\Delta t
\label{ti1}\\
t_{i+1}=t_s+i \Delta t
\label{ti2}
\een
Next, substituting (\ref{ti1}) and (\ref{ti2}) in (\ref{bart}) it follows that
\be
\bar{t}_i=t_s+i \Delta t -\frac{\Delta t}{2}
\label{barti}
\ee 
The mass of the spacecraft will decrease every time the thruster is fired to change the 
direction of the velocity at the times $t_i :  i=1 \dots N$. The amount of fuel mass that is burned
is itself dependent upon the current mass of the spacecraft. This mass of fuel burned at time $t_i$
may be written as (cf. \cite{curtis4})
\be
\Delta m_i=m_{ci}(1-e^{-\frac{\Delta v}{I_{sp}g_0}})
\label{deltami}
\ee
where $m_{ci}$ is the mass of the spacecraft at time $t_i$, $I_{sp}$ is the specific impulse of the spacecraft's
thruster, $\Delta v$ is the change in velocity required, and $g_0=9.81$ m$/s^2$ is the acceleration
due to gravity at the surface of Earth.

Thus for the mass of the spacecraft itself we have
\be
m_{c(i+1)}-m_{ci}=-\Delta m_i=-m_{ci}(1-e^{-\frac{\Delta v}{I_{sp}g_0}})
\ee
or
\be
m_{c(i+1)}=m_{ci}e^{-\frac{\Delta v}{I_{sp}g_0}}.
\ee
Denoting the initial mass of the spacecraft by $m_{c1}$ this leads to 

\be
m_{ci}=m_{c1}e^{-\frac{\Delta v}{I_{sp}g_0}(i-1)}
\label{mci}
\ee
Now using (\ref{mci}) in (\ref{ii2}), it follows from (\ref{barfi}) that
\be
\bar{F}_i=\frac{2Gm_a\sin\theta_b}{h\Delta t}m_{c1}e^{-\frac{\Delta v}{I_{sp}g_0}(i-1)}
\ee
or
\be
\bar{F}_i=\lambda m_{c1}e^{-\frac{\Delta v}{I_{sp}g_0}(i-1)}
\label{barf2}
\ee
where we have defined the constant $\lambda$ for brevity such that
\be
\lambda=\frac{2Gm_a\sin\theta_b}{h\Delta t}
\label{beta}
\ee
Substituting (\ref{barf2}) in (\ref{dzi}) and using
(\ref{barti}) we have
\be
\Delta \zeta_i=-\frac{\kappa}{m_a}(t_e-t_s-i\Delta t -\frac{\Delta
  t}{2})\lambda v_{ai} m_{c1}e^{-\frac{\Delta v}{I_{sp}g_0}(i-1)}\Delta t
\ee
or
\be
\Delta \zeta_i=-\frac{\kappa}{m_a}(t_e-t_s-\frac{\Delta
  t}{2})\lambda v_{ai}m_{c1}e^{-\frac{\Delta v}{I_{sp}g_0}(i-1)}\Delta t
+\frac{\kappa}{m_a}\Delta t^2 i \lambda v_{ai} m_{c1}e^{-\frac{\Delta v}{I_{sp}g_0}(i-1)}
\label{thisdzi}
\ee
As has already been discussed, the time interval $\Delta t$ for one
pass of the spacecraft on the orbit segment is small in relation to the
time interval $t_e-t_s$, i.e. how long before the projected collision
the gravity tractor action starts. We can therefore neglect the term
$\frac{\Delta t}{2}$ in the expression $(t_e-t_s-\frac{\Delta t}{2})$
in (\ref{thisdzi}). Now using (\ref{totaldeltaz}) we have
\be
\Delta \zeta= -\frac{\kappa}{m_a}(t_e-t_s)\lambda m_{c1}\Delta t \sum_{i=1}^Nv_{ai} e^{-\frac{\Delta v}{I_{sp}g_0}(i-1)}
+\frac{\kappa}{m_a}\Delta t^2 \lambda m_{c1}\sum_{i=1}^N v_{ai} i e^{-\frac{\Delta v}{I_{sp}g_0}(i-1)}
\label{sums}
\ee

Further evaluation (\ref{sums}) requires taking into account the slowly varying velocity $v_a$, unless the asteroid is on a circular orbit when $v_{ai}$ will
be constant. In that case, denoting the constant velocity of the
asteroid by $v_a$ and defining
\be
q=\frac{\Delta v}{I_{sp}g_0}
\label{q1}
\ee
for brevity, the sums in (\ref{sums}) may be evaluated to give (see for example \cite{grad})
\be
\Delta \zeta= -\frac{\kappa}{m_a}\lambda v_am_{c1}\left((t_e-t_s)\Delta t \frac{e^{q-Nq}(e^{Nq}-1)}{e^q-1}-\Delta t^2 \frac{e^{q-Nq}(e^{q+Nq}+N-e^q(1+N))}{(e^q-1)^2}\right)
\label{sumsq2}
\ee
An example of the evaluation of (\ref{sums}) on an elliptic orbit,
i.e. where the velocity of the asteroid shows large variations, will be
discussed in Section 6.

\section{Design of the orbit segment: maximizing the average force}
As follows from (\ref{deltazetafinal}) maximizing the average force
that is exerted on the asteroid in each time interval $[t_i,t_{i+1}]$
corresponds to the largest possible $\Delta \zeta$ that can be
obtained, with all other variables fixed. In what follows, we
will consider the various types of Keplerian orbits that the  
spacecraft may move along, and how the average force depends on their
characteristics. The aim is to choose these characteristics so that
the average force, and therefore the final deflection, are as large as
possible. 
The attainable average forces will be compared to the
constant forces that can be obtained from the stationary gravity
tractor and the displaced-orbit gravity tractor.
\subsection{Conditions for avoiding plume impingement}
It may be noted that based on (\ref{ii2}) alone, the overall impulse
 from one pass of the spacecraft on the orbit segment will
be largest if $h$ is chosen as small 
as possible and if $\theta_b=\frac{\pi}{2}$. Using (\ref{dzinew}) this
would also result in the largest possible contribution to the
deflection from that time interval. However, the need to avoid
impingement of the thruster plume on the asteroid will modify the
problem. Thus in this section we consider the problem of avoiding
plume impingement during restricted Keplerian motion. The problem is analogous to
the consideration of the exhaust plume that is done for the stationary
gravity tractor in \cite{lu-love}. 

\begin{figure}[htbp]
  \begin{center}
    \unitlength=.5in
    \begin{picture}(4,6.5)
\put(-2,-.5){\includegraphics[]{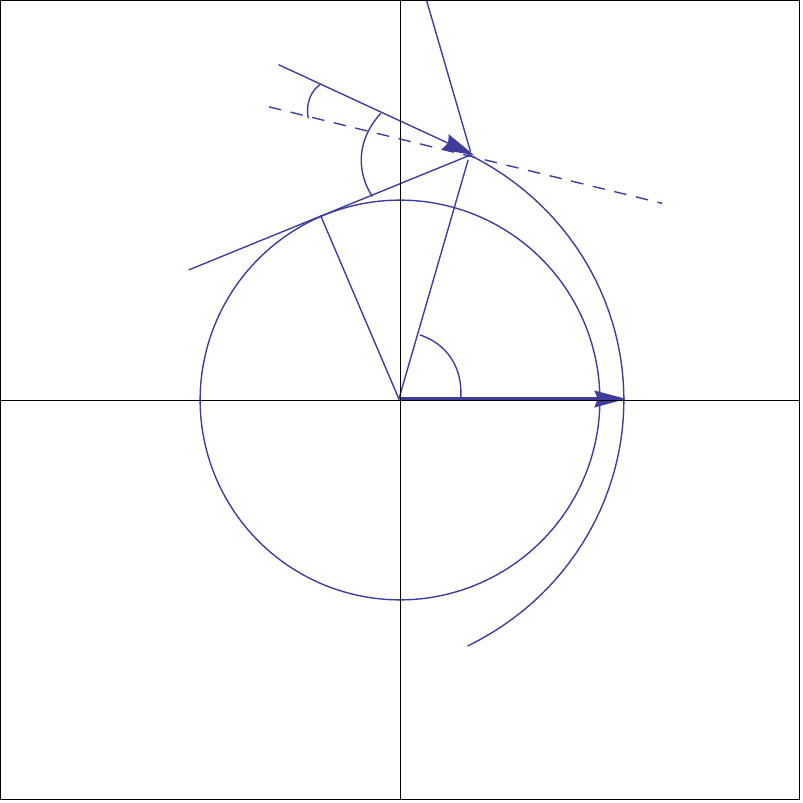}}
     \put(1.5,3.5){\small{$r(\theta_b)$}}
     \put(1.8,2.9){\small{$\theta_b$}}
     \put(1.8,2.5){\small{$r_{pm}$}}
     \put(.3,5.15){\small{$\gamma$}}
     \put(.3,5.45){\small{$F$}}
     \put(.6,4.5){\small{$\varphi$}}
     \put(.55,3.5){\small{$r_a$}}

    \end{picture}
  \end{center}
  \caption{Effect of plume angle on minimum periapsis radius.}
\label{plume}
\end{figure}
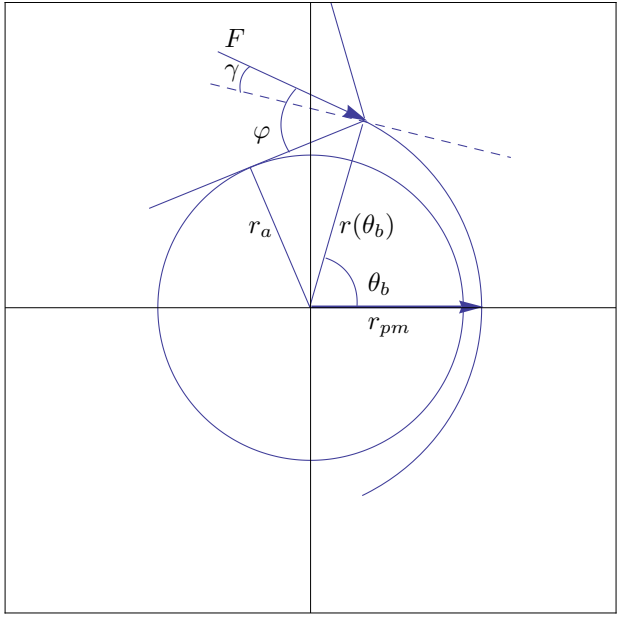

Figure \ref{plume} shows a situation with a spacecraft firing its
thruster to generate a force $F$. The plume half angle is denoted as $\varphi$,
and $\gamma$ is the flight path angle at that point on the orbit
segment given by (cf. \cite{curtis2})
\be
\gamma=\tan^{-1}\left(\frac{e \sin\theta_b}{1 + e \cos\theta_b}\right)\;.
\label{gamma}
\ee 

It is clear that plume impingement can be avoided by choosing a sufficiently large value for $r(\theta_b)$. Noting that $r(\theta_b)$ depends on the periapsis distance $r_p$, the eccentricity $e$, and the value of $\theta_b$,
a condition to avoid plume impingement can be derived as follows.

Referring to Figure \ref{plume} we note that
\be
r_a=r(\theta_b) \cos(\varphi-\gamma)\;,
\label{ra}
\ee
where $r_a$ is the radius of the asteroid (or in the case of a
non-spherical asteroid, the largest distance on the asteroid from the
center of mass in the plane of motion of the spacecraft). An expression for $r(\theta_b)$ can be found from the 
equation of path for the two-body problem (cf. \cite{curtis2,prussing1})
\be
r(\theta)=\frac{r_{p}(1+e)}{1+e\cos\theta}\;,
\label{eqp}
\ee
with $\theta=\theta_b$. Denoting the smallest allowable
value of $r_p$ by $r_{pm}$ and using (\ref{eqp}) and (\ref{ra}),
it follows that
\be
r_{pm}=\frac{(1+e\cos\theta_b)}{(1+e)\cos(\varphi-\gamma)}r_a\;.
\label{rpm}
\ee
Thus (\ref{rpm}) allows us to determine how close a given orbit
segment can come to the asteroid without violating the plume non-impingement
condition. This in turn determines the maximum average force that can
be obtained from the orbit segment as will be seen in the following sections.


\subsection{Restricted motion along a  circular orbit}
The simplest orbit segment that can be used to impart an impulse to the
asteroid per the discussion above is one that is circular.
In this case, $r_{pm}$ is independent of $\theta_b$ as follows from
(\ref{rpm}) by using $e=0$ and $\gamma=0$. The resulting value
\be
r_{pm}=\frac{r_a}{\cos\varphi}\;.
\label{rm}
\ee
is simply the radius of the smallest circle that satisfies the
non-impingement condition.

For a given value of the bounding angle $\theta_b$ the time of flight
between points $A$ and $B$ on a circular orbit of radius $r$ is (cf. \cite{curtis3,prussing2})
\be
\Delta t=2\theta_b\sqrt{\frac{r^3}{\mu}}\;.
\label{tcirc}
\ee
Thus considering a single pass on such an orbit segment with bounding angle
$\theta_b$ and using (\ref{impab4}) with $e=0$ and $r_p=r$, the net
impulse that is imparted to the asteroid per unit time (or average force as
defined in (\ref{barfi}) can be calculated as  
\be
\bar{F}_i=\frac{I_i}{\Delta t}=\frac{Gm_{ci}m_a\sin\theta_b}{\theta_b r^2}\;,
\label{etatd}
\ee
where as before $m_{ci}$ is the mass of the spacecraft during the
corresponding time interval $[t_i,t_{i+1}]$.
Using (\ref{rm}) to find the smallest allowable radius in order to
maximize the average force
\be
\bar{F}_{i(\mbox{max})}=\frac{Gm_{ci}m_a}{r_a^2}\frac{\sin\theta_b\cos^2\varphi}{\theta_b}\;.
\label{didt}
\ee
For the sake of comparison with the stationary and displaced-orbit gravity tractors,
we define the nondimensional maximum average force (for given values of
the plume angle $\varphi$ and the bounding angle $\theta_b$).
\be
\eta_k(\theta_b,\varphi)=\frac{\sin\theta_b\cos^2\varphi}{\theta_b}\;,
\label{etad}
\ee
by scaling (\ref{didt}) by $\frac{Gm_{ci}m_a}{r_a^2}$. 

\subsubsection{Comparison of the deflecting forces}
The nondimensional force for the stationary gravity tractor can be defined
the same way as in (\ref{etad}) and based on (\ref{fis}):
\be
\eta_s=\frac{1}{\alpha^2}\;.
\ee
The value for $\alpha$ 
in for example \cite{lu-love,olympio} is
$\alpha=1.5$. It has been shown, however, (see \cite{broschart}) that this value
(and small values of $\alpha$ in general) may lead to
oscillations of the altitude of the hovering  spacecraft with respect
to the asteroid.  Therefore, as in \cite{mcinnes2}, we also consider
$\alpha=2.5$ as a more realistic value for comparisons below. Overall, it may be noted that a small $\alpha$ leads to a
large cant angle and therefore inefficiency with respect to thrust
developed by the spacecraft. Similarly a large $\alpha$ means a
smaller force between the asteroid and the spacecraft lowering the
overall impulse imparted. 

Lastly for the nondimensional force from the displaced-orbit gravity tractor we have
\be
\eta_d=0.21
\ee
which follows by scaling (\ref{fdi}) by $\frac{Gm_{ci}m_a}{r_a^2}$. 

\begin{figure}[htbp]
  \begin{center}
    \unitlength=.5in
    \begin{picture}(4,5)
\put(-2,-.5){\includegraphics[]{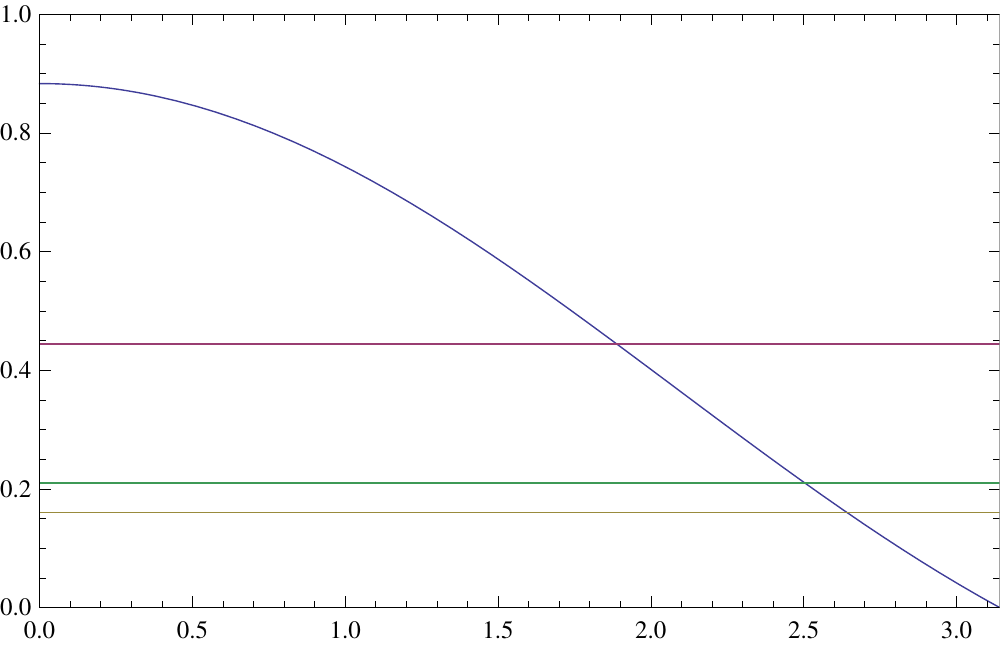}}
      \put(0,1.6){\small{$\eta_s(\alpha=1.5)$}}
      \put(0,.3){\small{$\eta_s(\alpha=2.5)$}}
      \put(0,.9){\small{$\eta_d=0.21$}}
      \put(-3.3,4.5){{$\eta_k(\theta_b,\varphi)$}}
      \put(6.3,-.4){\small{$\theta_b$}}
      \put(1.5,3){\small{$\eta_k(\theta_b,\varphi)$}}
    \end{picture}
  \end{center}
  \caption{A comparison of the average force exerted by the
    stationary, displaced-orbit, and Keplerian gravity tractor in a circular orbit ($\varphi=20^{o}$).}
\label{fig1}
\end{figure}
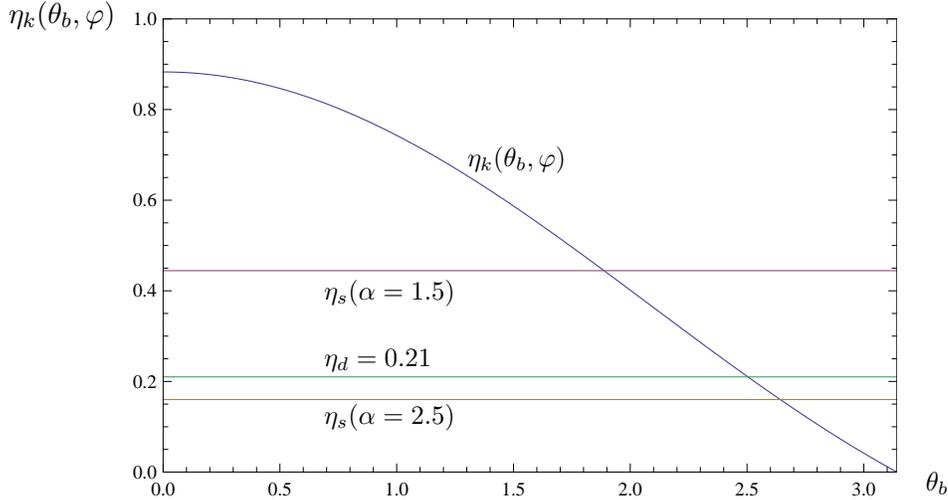

Figure \ref{fig1} shows the nondimensional average forces that are exerted by
the Keplerian (circular orbit), stationary, and displaced-orbit
gravity tractors. The average force from the Keplerian gravity tractor
depends on the value of the bounding angle $\theta_b$ and is plotted
as a function of $\theta_b$. The plume half angle
which also affects the net force from the Keplerian
gravity tractor is here set to $\varphi=20^{o}$ for the sake of numerical
comparison. The average force from the stationary and
displaced-orbit gravity tractors do not depend on $\theta_b$ and are
therefore shown as constant values. It follows that
for an 
appropriately chosen bounding angle, i.e. $\theta_b<1.9$ rad the Keplerian gravity tractor will
exert the largest net force on the asteroid. 

Overall, the average force from  the Keplerian
gravity tractor is large because of the closeness of the spacecraft to
the asteroid that can be achieved. The average force is especially large for small
values of $\theta_b$ (as the orbit segments remain close to periapsis). However, there will be limits on how small
$\theta_b$ can be made in practice. For example, when the burn times
required to effect the required changes in velocity of the spacecraft
are not much smaller than the time between burns, the present analysis
is not strictly valid. A general discussion of the time between burns is
presented in Section 4.4

\subsection{Restricted motion along non-circular orbits}

The time of flight (TOF) between the angles $\theta=-\theta_b$ and
$\theta=\theta_b$ on an orbit of eccentricity $e$ and periapsis
distance $r_p$ is determined as (see for example \cite{curtis3,prussing2})
\ben
\Delta t_{e}&=&2\sqrt{\frac{r_p^3}{\mu_a(1-e)^3}}(E_b-e\sin
E_b)\;\;\;\;\;e<1\;\; (\mbox{ellipse})\;,
\label{te}\\
\Delta t_{h}&=&2\sqrt{\frac{-r_p^3}{\mu_a(1-e)^3}}(F_b-e\sin
F_b)\;\;\;\;\;e>1\;\; (\mbox{hyperbola})\;,
\label{th}\\
\Delta t_{p}&=&\left(\frac{(2
  r_p)^{3/2}}{\sqrt{\mu_a}}\right)(\frac{1}{3}\tan\frac{\theta_b^3}{2} +
\tan\frac{\theta_b}{2})\;\;\;\;\;e=1\;\; (\mbox{parabola})\;,
\label{tp}
\een
where $E_b$ is the eccentric anomaly determined by
\be
\cos E_b=\frac{e+\cos\theta_b}{1+e\cos\theta_b}\;\;\;\;\; (e<1)\;,
\ee
and $F_b$ is the hyperbolic anomaly determined by
\be
\cosh F_b=\frac{e+\cos\theta_b}{1+e\cos\theta_b}\;\;\;\;\; (e>1)\;.
\ee
Thus using (\ref{impab4}) the impulse imparted per unit time (or average force) for the elliptic orbit can be
written as
\be
\frac{I_i}{\Delta t_{e}}=\frac{Gm_am_{ci}}{r_p^2}\sqrt{\frac{(1-e)^3}{1+e}}\frac{\sin\theta_b}{E_b-e\sin
  E_b}\;,
\label{didte}
\ee
and for the parabolic orbit
\be
\frac{I_i}{\Delta t_{p}}=\frac{Gm_am_{ci}}{4
  r_p^2}(\frac{1}{3}\tan^3\frac{\theta_b}{2}+\tan\frac{\theta_b}{2})\;.
\label{didtp}
\ee
Similarly, for the hyperbolic orbit, it is obtained that
\be
\frac{I_i}{\Delta
  t_{h}}=\frac{Gm_am_{ci}}{r_p^2}\sqrt{\frac{(e-1)^3}{1+e}}\frac{\sin\theta_b}{e\sin
  F_b-F_b}\;.
\label{didth}
\ee
In each case, the largest impulse per time is obtained
if $r_p$ is chosen as $r_{pm}(\theta_b)$ given by (\ref{rpm}). Thus,
using (\ref{rpm}) in (\ref{didte})-(\ref{didth}) and scaling each
average force by $\frac{Gm_am_{ci}}{r_a^2}$ we define the
nondimensional average force $\eta(\theta_b,\varphi,e)$ through
\be
\eta(\theta_b,\varphi,e)=\left\{\begin{array}{l}
\frac{1}{A(\theta_b,\varphi,e)}\sqrt{\frac{(1-e)^3}{1+e}}\frac{\sin\theta_b}{E_b-e\sin
  E_b}\;\; (e<1)\;,\\
\\
\frac{1}{4A(\theta_b,\varphi,e)}(\frac{1}{3}\tan^3\frac{\theta_b}{2}+\tan\frac{\theta_b}{2})\;\;
(e=1)\;,\\
\\
\frac{1}{A(\theta_b,\varphi,e)}\sqrt{\frac{(1-e)^3}{1+e}}\sqrt{\frac{(e-1)^3}{1+e}}\frac{\sin\theta_b}{e\sin
  F_b-F_b}\;\; (e>1)\;,
\end{array}\right.
\ee
where, using (\ref{rpm}),  we have defined
\be
A(\theta_b,\varphi,e)=\frac{(1+e\cos\theta_b)^2}{(1+e)^2\cos^2(\varphi-\gamma)}
\ee
for brevity.

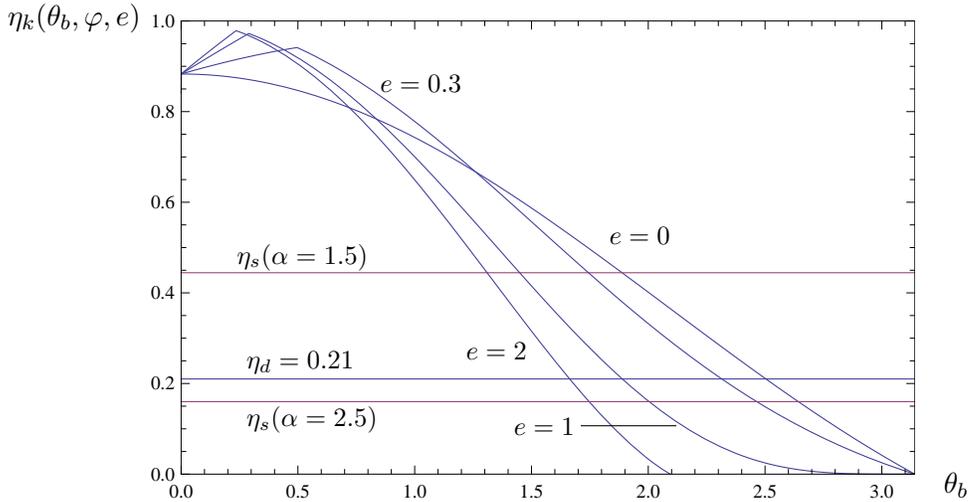
\begin{figure}[htbp]
  \begin{center}
    \unitlength=.5in
    \begin{picture}(4,5)
\put(-2,-.5){\includegraphics[]{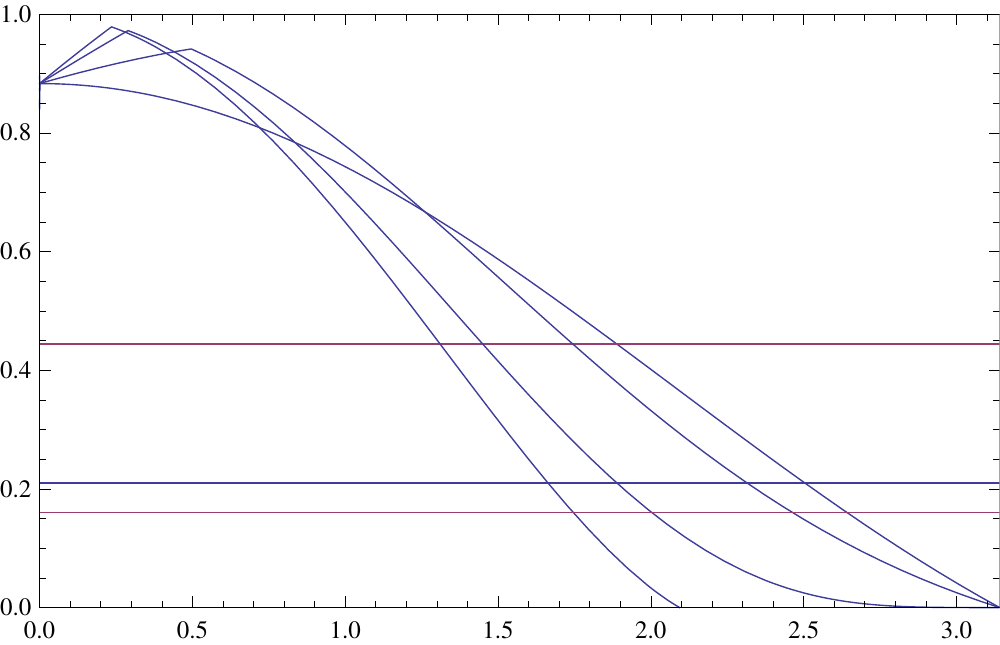}}
\put(-1.1,2){\small{$\eta_s(\alpha=1.5)$}}
\put(1.3,1){\small{${e=2}$}}
\put(1.8,0.2){\small{$e=1$}}
\put(2.5,0.3){\line(1,0){1}}
\put(-3.5,4.5){$\eta_k(\theta_b,\varphi,e)$}
\put(6.3,-.4){$\theta_b$}
\put(.4,3.8){\small{$e=0.3$}}
\put(2.8,2.2){\small{$e=0$}}
      \put(-1,.3){\small{$\eta_s(\alpha=2.5)$}}
      \put(-1,.9){\small{$\eta_d=0.21$}}
    \end{picture}
  \end{center}
  \caption{A comparison of the average force exerted by the Keplerian
    gravity tractor on orbits of various eccentricities. The
    corresponding forces from the stationary gravity tractor and the
    displaced-orbit gravity tractor are also shown for comparison. The
  plume angle is $\varphi=20^o$.}
\label{conics}
\end{figure}

Thus, as in the case of the circular orbit, the average
force exerted on the asteroid can be studied as a function of
$\theta_b$, now with the
eccentricity $e$ acting as a parameter modifying the orbit. 
Figure \ref{conics} shows plots of $\eta_k(\theta_b,\varphi,e)$ with
respect to $\theta_b$ and for various values of the eccentricity
$e$. The value of $\varphi$ used is again $20^{o}$. For comparison,
the earlier derived corresponding values of $\eta_s$ and $\eta_d$ are
also shown. 
We conclude from Figure \ref{conics} that even though according to (\ref{impab2}) alone,
the best solution would have been to always choose $e=0$, the plume
non-impingement condition means that there are values of $\theta_b$
for which a larger average force can be obtained with $e\neq0$. The
overall largest values of $\eta_k$ are obtained for small values of
$\theta_b$ and large values of the eccentricity. It should be noted,
however, that these cases 
may be difficult to realize in practice due to the corresponding short
times between impulsive thrusts. In addition, as in the case of the
circular orbit we not that when the flight times
approach the thrust durations necessary to affect the required $\Delta
v$'s the assumption of instantaneous impulsive thrust will not be
satisfied and further analysis would be necessary. This will be
further discussed in the next section.

\subsection{The time between impulsive thrusts}
As is common in the preliminary analysis of impulsive velocity
changes ($\Delta v$'s), we have above assumed that 
the impulsive thrusts that change the direction of the
velocity of the spacecraft moving along a Keplerian orbit segment,
are applied instantaneously. In practice, of course, the
implication is only that the duration of the thrusts is small compared
to a characteristic time of flight on the orbit. For this reason, it
is important to consider the time of flight on the Keplerian
orbit segment and verify that it is much larger than the duration of a
typical thrust.

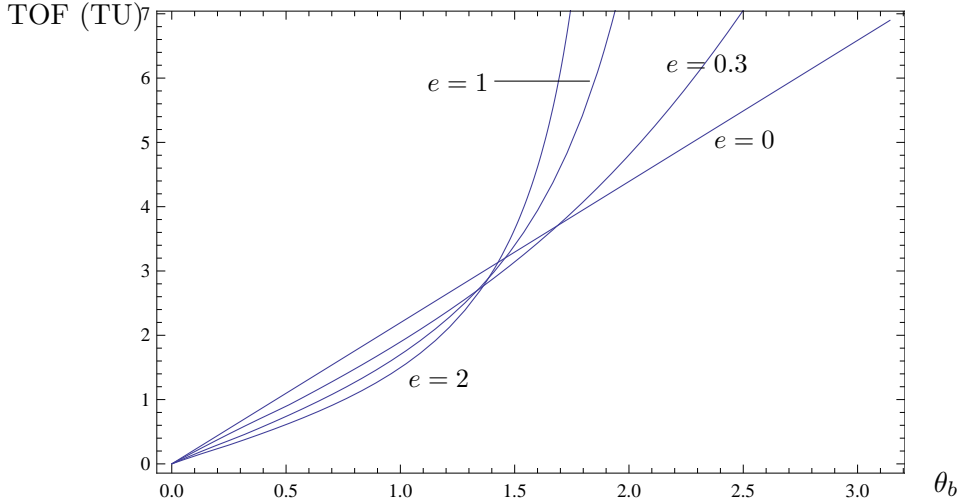
\begin{figure}[htbp]
  \begin{center}
    \unitlength=.5in
    \begin{picture}(4,5)
\put(-2,-.5){\includegraphics[]{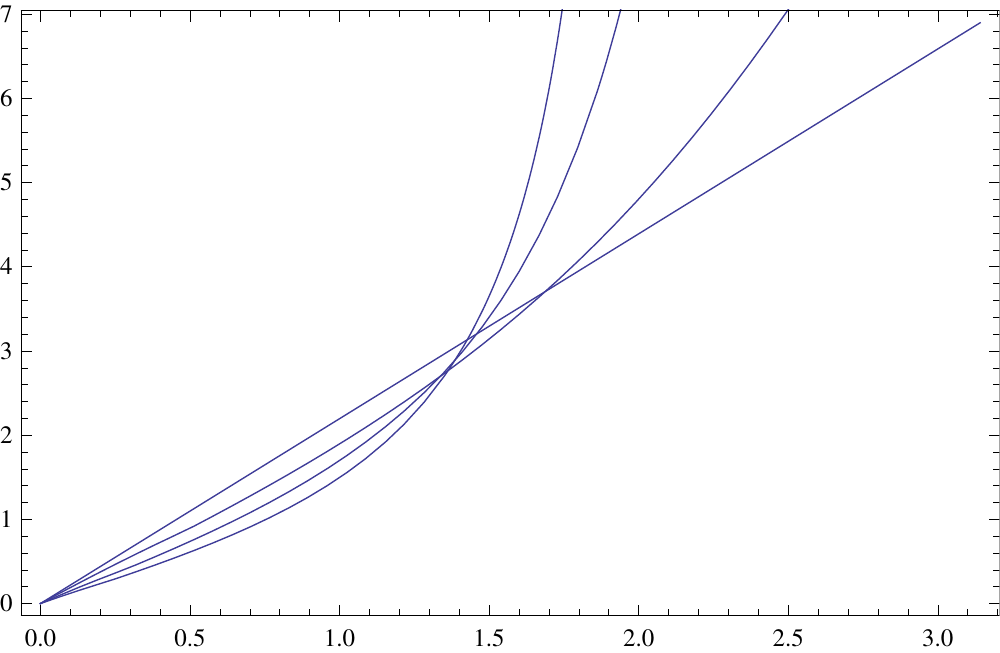}}
\put(-3.4,4.5){TOF (TU)}
\put(.8,.7){\small{${e=2}$}}
\put(1,3.8){\small{$e=1$}}
\put(1.7,3.9){\line(1,0){1}}
\put(6.3,-.4){$\theta_b$}
\put(3.5,4){\small{$e=0.3$}}
\put(4.0,3.2){\small{$e=0$}}
    \end{picture}
  \end{center}
  \caption{Times between impulsive thrusts as a function of $\theta_b$
    on the orbit segments for various eccentricities; $\varphi=20^{o}$.}
\label{times}
\end{figure}

The time between the impulsive thrusts, i.e. between the angles
$-\theta_b$ and $\theta_b$, or vice-versa, can be obtained directly from
(\ref{te})-(\ref{tp}), where $r_p$ is set to $r_{pm}$ using
(\ref{rpm}). The resulting times of flight are shown in Figure
\ref{times}. Canonical 
time units for the asteroid are used such that (cf. \cite{bmw})
\be
1 \mbox{ TU} = \sqrt{\frac{r_a^3}{\mu_a}}\;.
\ee
To get a sense of the length of 1~TU, consider as before an asteroid
of diameter 100~m and mass $8.4\times 10^{9}$~kg. Then, it follows that  1~TU$ \approx 1340 s=22$ min. 
Therefore, for example, near the value of $\theta_b=1.5$ rad where, for small
eccentricities, $\eta_k(\theta_b,\varphi,e)\approx 0.6$ (see Figure \ref{conics})
the time between impulsive thrusts is in the order of 3 TU or 66
minutes.

One implication of the time between impulsive thrusts is that the
actual duration of the thrusts themselves must be much smaller than
the time between the thrusts. In the next section, we derive an
expression for the required $\Delta v$'s and estimate the
corresponding thrust times in order to verify that the assumption of
impulsive thrusts is valid.

\subsection{The impulsive velocity change}
In order to change the direction of the velocity of
the spacecraft on the orbit segment, the required $\Delta v$ is exactly twice
the magnitude of the velocity. The velocity at the bounding angles 
can be obtained from the expression for the (specific) energy of the orbit
(cf. \cite{curtis2,prussing1})
\be
{\cal E}=\frac{v^2}{2}-\frac{\mu_a}{r}=-\frac{\mu_a}{2a}\;,
\ee
where $a$ is the semimajor axis of the orbit. Noting that the
semimajor axis can be written as (cf. \cite{curtis2,prussing1})
\be
a=\frac{r_p}{1-e}\;,
\ee
and using the equation of path (\ref{eqp}), it follows that the
velocity magnitude at the bounding angles is 
\be
v_m=\sqrt{\frac{\mu_a}{r_p(1+e)}}\sqrt{1+e^2+2e\cos\theta_b}\;.
\ee
Then, the $\Delta v$ on the orbit with $r_p=r_{pm}$ is
\be
\Delta v=2 \sqrt{\frac{\mu_a}{r_{pm}(1+e)}}\sqrt{1+e^2+2e\cos\theta_b}\;,
\label{dv}
\ee
or using (\ref{rpm}) in (\ref{dv})
\be
\Delta v=2 \sqrt{\frac{\mu_a}{r_{a}}}\sqrt{\frac{\cos(\varphi-\gamma)}{1+e\cos\theta_b}}\sqrt{1+e^2+2e\cos\theta_b}\;.
\label{dv2}
\ee
We note that the expression for $\Delta v$ in (\ref{dv2}) is in the form
\be
\Delta v= \sqrt{\frac{\mu_a}{r_{a}}}\nu(\theta_b,e,\varphi)\;,
\ee
where we have defined 
\be
\nu(\theta_b,e,\varphi)=2\sqrt{\frac{\cos(\varphi-\gamma)}{1+e\cos\theta_b}}\sqrt{1+e^2+2e\cos\theta_b}\;.
\ee
and $\sqrt{\frac{\mu_a}{r_{a}}}$ is the circular satellite velocity at a
radius of $r_a$ around the asteroid.

\begin{figure}[htbp]
  \begin{center}
    \unitlength=.5in
    \begin{picture}(4,5)
\put(-2,-.5){\includegraphics[]{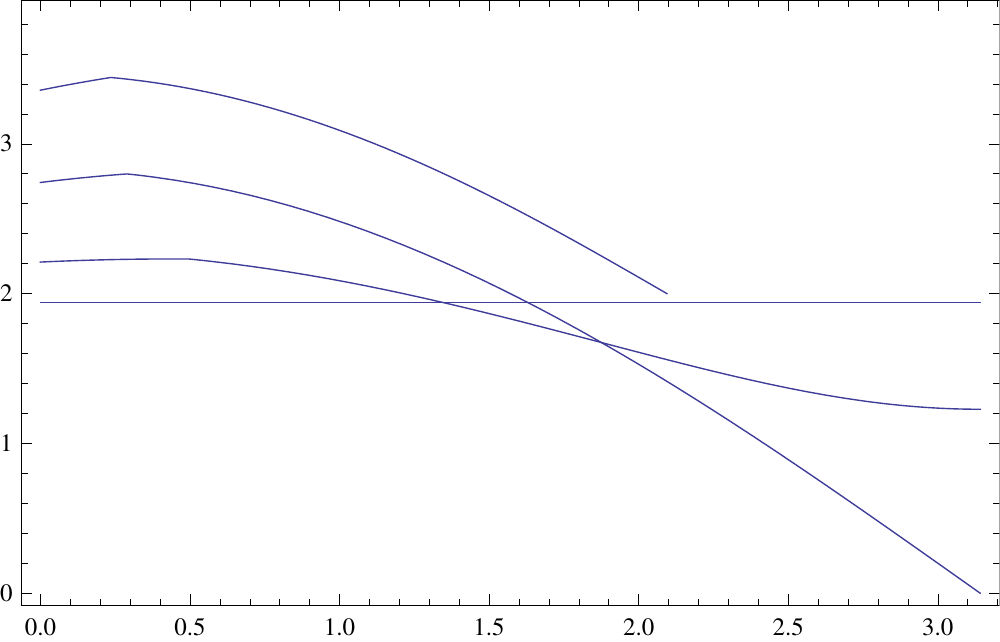}}
\put(-3.4,4.5){$\nu(\theta_b,e,\varphi)$}
\put(4.8,.7){\small{${e=1}$}}
\put(.6,3.7){\small{$e=2$}}
\put(6.3,-.4){$\theta_b$}
\put(-1,2){\small{$e=0$}}
\put(4.25,1.6){\small{$e=.3$}}
    \end{picture}
  \end{center}
  \caption{Nondimensional impulsive $\Delta v$'s as functions of the
    bounding angle $\theta_b$ for various values of $e$; $\varphi=20^{o}$.}
\label{deltavs}
\end{figure}
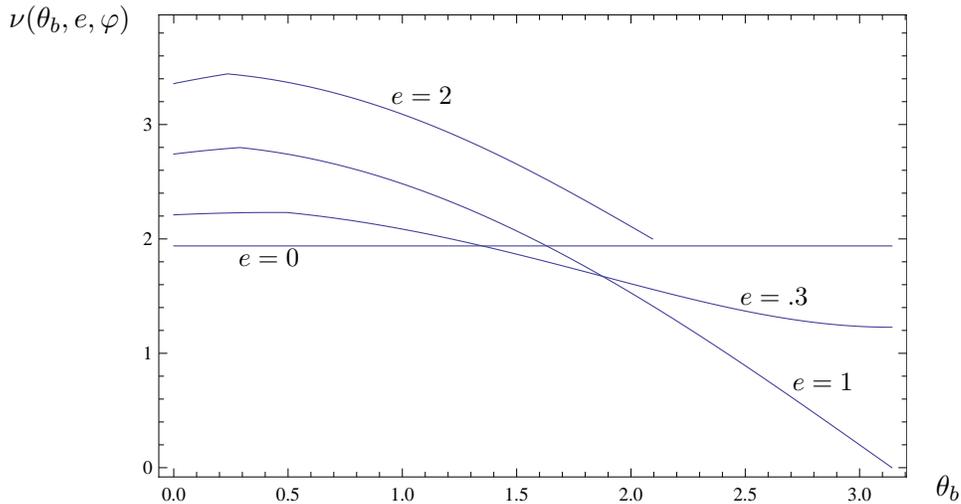

Figure \ref{deltavs} shows the dependence of $\nu(\theta_b,e,\varphi)$
on $\theta_b$ and $e$. The value of $\varphi=20^{o}$ has been used for
the sake of example. 
Thus for an orbit of given eccentricity, the required $\Delta v$ will
be smallest for large bounding angles as the velocity will decrease
with increasing true anomaly. The
required $\Delta v$ generally increases with decreasing bounding
angle. However, for small values of the bounding angle and a given
eccentricity, 
it is seen that there is a region where the required $\Delta v$
decreases with the bounding angle $\theta_b$. This is due to the fact
that for these 
values of $\theta_b$ the spacecraft is generally close to the surface
of the asteroid and the plume impingement constraint is
active. Consequently, as $\theta_b$ decreases, the minimum periapsis
distance is required to increase leading to a smaller velocity $v_b$ and therefore a
smaller $\Delta v$.
  
The values of the required $\Delta v$'s have implications on the
choice of thrusters for the spacecraft. 
As an example we consider an asteroid of mass $8.4\times 10^9$ kg and
radius $100$ m (as before). The $\Delta v$ that would be required on a
circular orbit segment (with $\varphi=20^{o}$) is $0.145$ m/s. Now,
the impulsive thrust $F$ required to obtain a given
$\Delta v$ for the spacecraft can be found from the relation
\be
T\Delta t=m_c\Delta v \;.
\label{mom}
\ee
where $\Delta t$ is the duration of the thrust. If we assume, for the
sake of example, a spacecraft of mass 2000 kg, the impulse required is
290 N-s which could for example be obtained by a 1-N force acting over
290 seconds. The duration of 290 s can be compared to
the time of flight of 4600 s (that is obtained using (\ref{tcirc})
with $r_p$ set to $r_m$ from (\ref{rm}))
thus justifying the model of impulsive velocity changes.

\subsection{Full orbits between impulsive thrusts}
Depending on the configuration of the thrusters on a spacecraft, the
spacecraft may need some minimum amount of time in order to re-orient
itself between successive firings of the thrusters, i.e. so that the
thrusters are pointing in a desired direction.
In the case of circular and elliptic orbits, a technique to increase 
the time between impulsive thrusts, if necessary,  may be for the
spacecraft to complete one or more full orbits around the asteroid
between the bounding angles where the impulsive thrusts are
executed. For example, referring to Figure \ref{system} the spacecraft
could start at point $A$, move past point $B$ without an impulsive
thrust at that point, and return to point $B$ after a full cycle on
the orbit, at which time an impulsive thrust could change the
direction of motion sending the spacecraft back to $A$ in a clockwise
direction. In this way, while the impulsive thrusts required at $A$
and $B$ remain unchanged, as well as the net impulse on the asteroid, the time between the impulsive thrusts can
be increased. 

\begin{figure}[htbp]
  \begin{center}
    \unitlength=.5in
    \begin{picture}(4,5)
\put(-1.9,-.5){\includegraphics[]{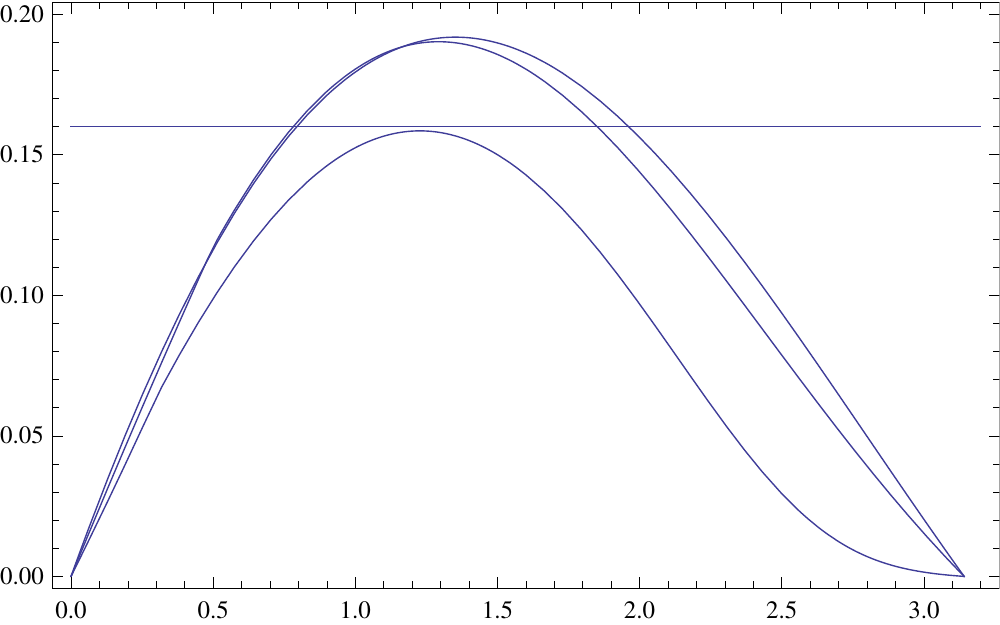}}
      \put(-4.2,4.4){$\eta_k(\theta_b+2\pi,e,\varphi)$}
      \put(6.3,-.4){$\theta_b$}
\put(1.8,2.5){\small{$e=0.9$}}
\put(6.3,-.4){$\theta_b$}
\put(3.7,3.8){\small{$e=0.3$}}
\put(3.6,3.8){\line(-1,-1){.53}}
      \put(-1.3,3.6){\small{$\eta_s(\alpha=2.5)$}}
\put(4.1,2.5){\small{$e=0$}}
    \end{picture}
  \end{center}
  \caption{Impulse per time imparted by elliptic orbits with one extra
orbit cycle; $\varphi=20^{o}$}
\label{cycles}
\end{figure}
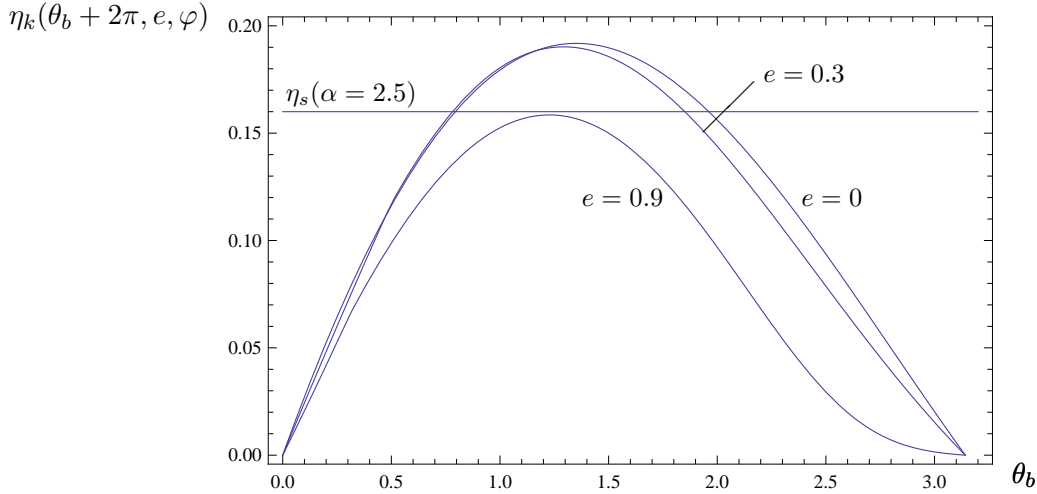

Figure \ref{cycles} shows the average force for the case of
an added orbit period, as a function of the
bounding angles, for three values of the eccentricity. For comparison,
the average force from the stationary gravity tractor with $\alpha=2.5$ is
also shown. 

Overall, as would be expected, the impulse per time is smaller for all angles
of $\theta_b$ in comparison to the case of no extra orbit cycles.
This decrease is especially large for small values of
$\theta_b$. However, this approach can still give a larger average force
than the stationary gravity tractor with $\alpha=2.5$.

\section{Impulse imparted per mass of fuel used}

The contribution to the total
deflection at the time of encounter $t_e$ from an impulse imparted to
the asteroid at time $t_i$ is given in (\ref{dzinew}). Thus
$\Delta\zeta_i$ in (\ref{dzinew}) is proportional to the impulse
$I_i$. Therefore a consideration of impulse imparted $I_i$ 
per amount of fuel burned in the
spacecraft during that same time interval gives another measure of the
efficiency of the different asteroid deflection methods.
In this section we derive expressions for the impulse imparted per
mass of fuel burned and compare them between
the Keplerian, stationary, and displaced-orbit gravity tractors.

The mass of fuel burned during an
impulsive velocity change when the spacecraft mass is $m_{ci}$ is
given in (\ref{deltami}).
On the other hand, an expression for the impulse imparted to the
asteroid from one pass of the spacecraft on a Keplerian orbit segment
is given in (\ref{impab4}). Noting that an impulsive thrust takes
place at the beginning of each pass on a Keplerian orbit segment, the
impulse imparted during a time interval $[t_i,t_{i+1}]$ per mass of
fuel burned at the beginning of the interval can be expressed as

\be
\frac{I_i}{\Delta m_i}=\frac{2Gm_a\sin{\theta_b}}{\sqrt{\mu_ar_p(1+e)}(1-e^{-\frac{\Delta v}{I_{sp}g_0}})}\;.
\label{etamd}
\ee
This quantity will be referred to below as the {\em mass efficiency}.

Recall that a typical value for a $\Delta v$ is in the order of
0.1~m/s per the discussion just before (\ref{mom}). At the same time, 
$I_{sp}$ values realizable with current technology may range from
hundreds of seconds for chemical 
rockets to several 1000 seconds for ion engines (\cite{curtis4}).
Therefore it is clear that $I_{sp}g_0>>\Delta v$ and the ratio
$\frac{\Delta v}{I_{sp}g_0}$ in (\ref{etamd}) is generally small.
Using (\ref{dv}) useful approximation for the mass efficiency can
therefore be obtained from (\ref{etamd}) in the form
\be
\frac{I_i}{\Delta m_i}\approx \frac{I_{sp}g_0\sin\theta_b}{\sqrt{1+e^2+2e\cos\theta_b}}\;,
\label{etamd2}
\ee
meaning that it does not depend on the value of $r_p$. For easy
comparisons below, 
we define the nondimensional mass efficiency  by scaling the right hand side of (\ref{etamd2})
by $I_{sp}g_0$ so that
\be
\zeta_k(\theta_b,e)=\frac{\sin\theta_b}{\sqrt{1+e^2+2e\cos\theta_b}}\;,
\label{zetak}
\ee

\begin{figure}[htbp]
  \begin{center}
    \unitlength=.5in
    \begin{picture}(4,4)
     \put(2.8,2.2){$\beta+\varphi$}
     \put(-2,-1){\includegraphics[scale=0.4]{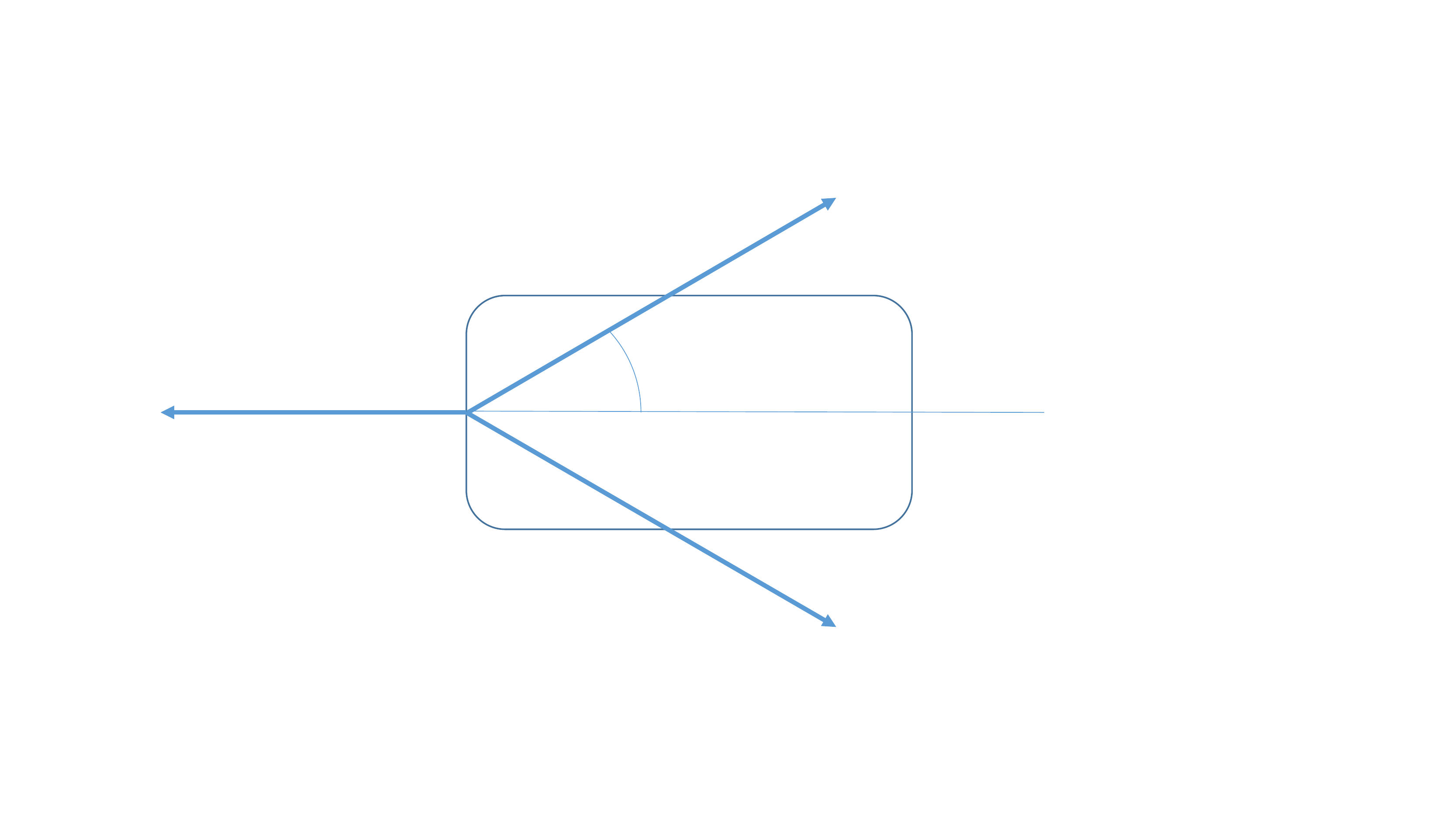}}
     \put(-1.2,2){$F_s$}
     \put(4.8,2.2){spacecraft}
     \put(3.6,3.6){$T$}
     \put(3.6,0.2){$T$}

    \end{picture}
  \end{center}
  \caption{Free body diagram for the stationary gravity tractor.}
\label{fig:fbd}
\end{figure}
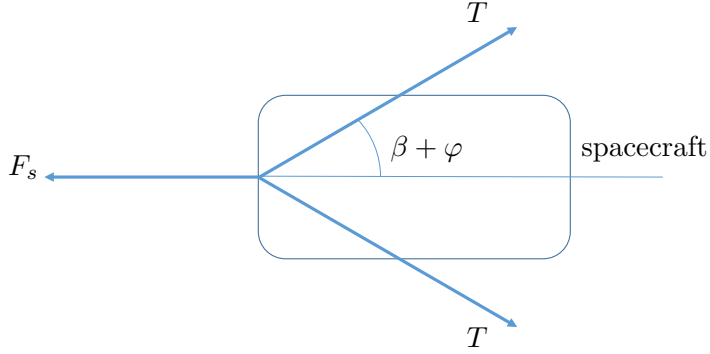

The corresponding mass efficiency for the stationary gravity tractor
can be found by using (\ref{etats}) and the free body diagram for the
spacecraft in Figure \ref{fig:fbd} to obtain a relation for the
required thrust, which leads to
\be
F_s=2T\cos(\beta+\theta)=\frac{Gm_am_c}{\alpha^2 r_a^2}\;.
\label{FT}
\ee

On the other hand, with respect to fuel expenditure, the thrust
required may be written in terms of the specific impulse 
and the rate of change of mass through (cf. \cite{curtis4})
\be
T=I_{sp}\frac{dm'}{dt}g_0\;.
\label{tisp}
\ee
where $dm'$ is an infinitesimal mass expended in {\em one} of the
thrusters in an infinitesimal time interval $dt$.
Thus the impulse imparted in a time interval $dt$ per mass of total fuel
expended during that time becomes (noting that $2dm'=dm$, i.e. the
change in mass of the spacecraft is twice the change in mass in each thruster)
\be
\frac{dI}{dm}=\frac{F_sdt}{2dm'}\;,
\ee
or using (\ref{FT}) and (\ref{tisp})
\be
\frac{dI}{dm}=I_{sp}g_0\cos(\beta+\theta);,
\label{dixdm}
\ee
indicating that the mass efficiency for the stationary gravity tractor
is constant and only depends on the specific impulse.
Scaling (\ref{dixdm}) by $I_{sp}g_0$ for a nondimensional mass
efficiency we obtain
\be
\zeta_s=\cos(\beta+\varphi)\;,
\ee
to be compared to $\zeta_k(\theta_b,e)$.

In the case of the displaced-orbit approach (see \cite{mcinnes2}), provided
that the condition is met for plume non-impingement, the force that is
acting on the asteroid is equal to the force that is developed by the
thruster, as follows from (\ref{tee}). Therefore
\be
\frac{dI}{dm}=\frac{F_ddt}{dm}=\frac{Tdt}{dm}
\ee
or using (\ref{tisp}) and noting that here $dm=dm'$
\be
\frac{dI}{dm}=I_{sp}g_0
\ee
The corresponding nondimensional mass
efficiency is therefore $\zeta_d=1$.

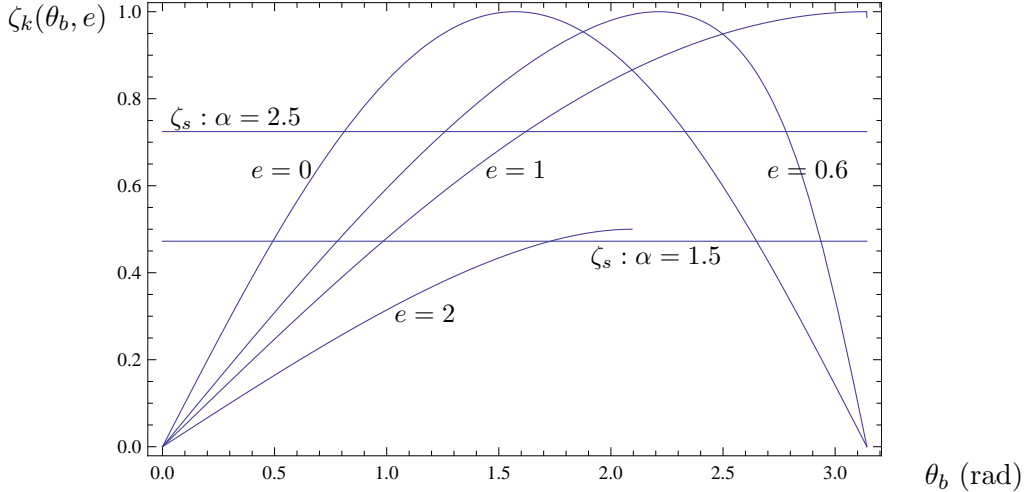
\begin{figure}[htbp]
  \begin{center}
    \unitlength=.5in
    \begin{picture}(4,5)
    \put(-2.15,-.2){\includegraphics[scale=1]{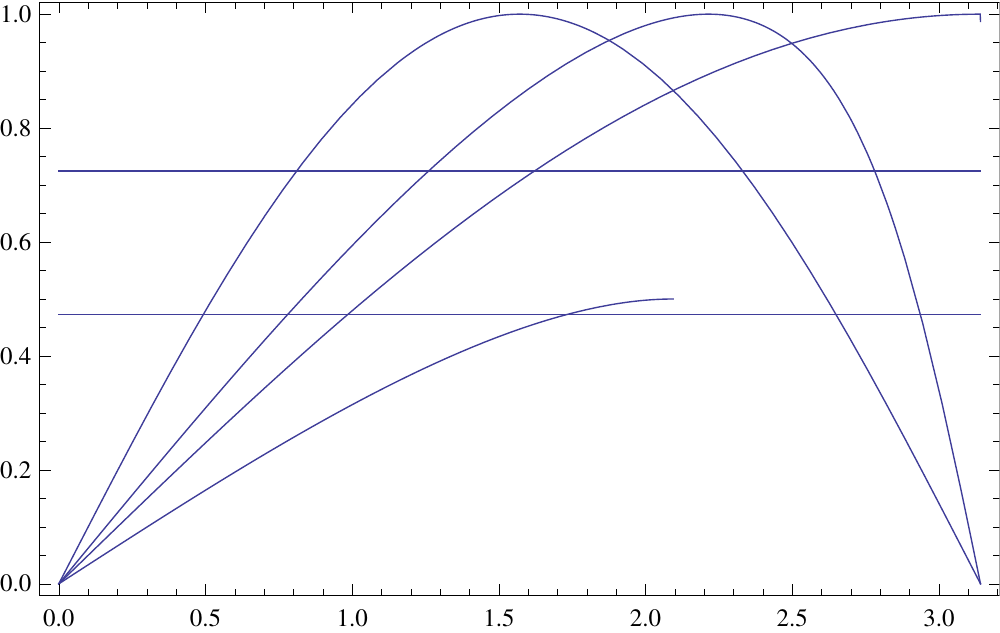}}
      \put(6.3,-.2){$\theta_b$ (rad)}
      \put(-3.3,4.6){$\zeta_k(\theta_b,e)$}
      \put(-.75,3){\small ${ e=0}$}
      \put(.75,1.5){\small ${ e=2}$}
      \put(1.7,3){\small ${ e=1}$}
      \put(4.65,3){\small ${ e=0.6}$}
      \put(2.8,2.1){\small ${ \zeta_s:\alpha=1.5}$}
      \put(-1.6,3.56){\small ${ \zeta_s:\alpha=2.5}$}
    \end{picture}
  \end{center}
  \caption{Comparison of impulse imparted per mass of fuel burned for
    the stationary and Keplerian gravity tractors.}
\label{etamh}
\end{figure}

Figure \ref{etamh} shows $\zeta_k(\theta_b,e)$ as a function of the
bounding angle $\theta_b$ for various values of the eccentricity
$e$. For the sake of comparison
$\zeta_s$ (for the values of $\alpha=1.5$ and $\alpha=2.5$) are shown the same diagram. 
The values of $\beta$ (see Figure \ref{fig:stat}) used for the stationary
gravity tractor corresponding to $\alpha=1.5$ and $\alpha=2.5$ and are found from
\be
\beta=\sin^{-1}\left(\frac{1}{\alpha}\right) \;,
\ee
which yields $\beta=41.8^{o}$ and $\beta=23.6^{o}$, respectively. 
In all cases, the plume angle is taken to be $\varphi=20^o$ as before.

Thus, for example, if one chooses $\theta_b=2.2$ rad and $e=0.6$ the
nondimensional mass efficiency for the Keplerian gravity tractor is
unity. At the same time, note that for the stationary gravity tractor
the mass efficiency $\zeta_s= 0.72$ (for $\alpha=2.5$) meaning that the
Keplerian gravity tractor has a 40\% higher mass efficiency in this
case. Similarly, for the case $\alpha=1.5$ ($\zeta_s=0.47$), the
Keplerian gravity tractor has a mass efficiency that is twice as large
as for the stationary gravity tractor.

Lastly, we note that adding full cycles of elliptic orbits
between the bounding angles $\pm \theta_b$, as detailed in Section
3.4, does not change the fuel 
requirements as the velocities at the points of impulsive thrusts
remain the same.

\section{Numerical Example}

In Section 4 we compared the average forces from the stationary, displaced-orbit, 
and Keplerian gravity tractors. Using (\ref{dzi}) and (\ref{eval3}) this allows for a 
direct comparison of the contribution to the deflection caused by these gravity tractors
from action during a small time interval $[t_i,t_{i+1}]$. In this section we consider a
more complete comparison of the Keplerian and stationary gravity tractors by considering the
total deflection caused by two similar spacecraft deployed as stationary and Keplerian gravity
tractors.

\subsection{An asteroid in an elliptic orbit: 2007 VK184}
As an example we consider the hypothetical deflection of asteroid 2007 VK184
that is projected to have a close encounter with Earth in June
2048, but with no risk of a collision. The orbital elements of the asteroid are given in Table \ref{table:vk184}, \cite{elements}. In addition, the mass and radius of the asteroid are $m_a=3.3\times10^9$ kg, and $r_a=65$ m, respectively, \cite{massradius}.

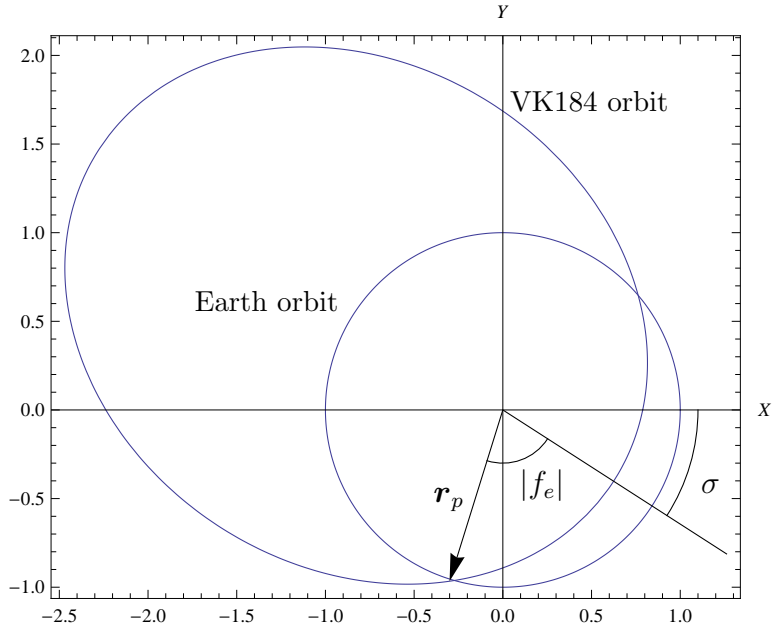
\begin{figure}[htbp]
  \begin{center}
    \unitlength=.5in
    \begin{picture}(4,7)
    \put(-2.15,-.2){\includegraphics[scale=1]{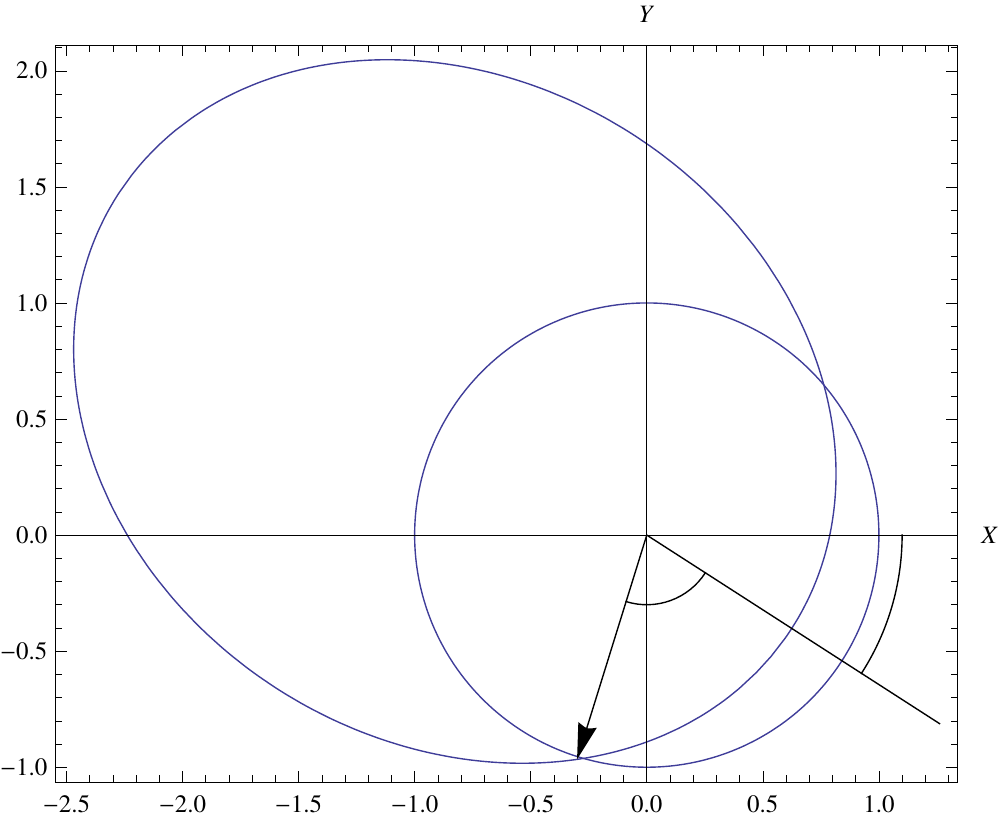}}
    \put(3.2,1.2){\large{$|f_e|$}}
    \put(5.1,1.2){\large{$\sigma$}}
    \put(3.1,5.2){{VK184 orbit}}
    \put(-.2,3.1){{Earth orbit}}
    \put(2.3,1.1){\large{$\br_p$}}
    \end{picture}
  \end{center}
  \caption{Heliocentric orbits of Earth and Asteroid 2007 VK184.}
\label{earth-vk184}
\end{figure}

\begin{table}[ht]
\caption{Orbital parameters for asteroid 2007 VK184} 
\centering 
\begin{tabular}{l c} 
\hline\hline 
Epoch & JD 2457600.5 (2016-07-31)\\ [0.5ex] 
\hline 
Semimajor axis & 1.7262 AU\\ 
Eccentricity & 0.5697\\
Mean Anomaly & 325.9$^{o}$\\
Inclination & 1.2209$^{o}$\\
Longitude of ascending node&253.64$^{o}$\\
Argument of perihelion&73.58$^{o}$\\ [1ex] 
\hline 
\end{tabular}
\label{table:vk184} 
\end{table}
For our purposes of comparing the efficiencies of the Keplerian and
stationary gravity tractors, we will neglect the small inclination of
the asteroid's orbit. Denoting the longitude of the ascending node by $\Omega$ and the argument of periapsis by $\omega$, the longitude of perihelion $\Pi$ may then be written as (cf. \cite{hale})
\be
\Pi=\Omega+\omega
\ee
or $\Pi=327.2^{o}$. Based on these assumptions, the orbits of the
asteroid and that of Earth may be depicted as in Figure  
\ref{earth-vk184} where Earth's orbit is taken to be circular 
(see also \cite{olympio}). The angle $\sigma$ in the figure is
\be
\sigma=2\pi-\Pi
\ee
and has the value of $\sigma=0.572$ rad, and the position of the
projected encounter is shown by the vector $\br_p$.
Denoting the true anomaly on the asteroid's orbit at the time of
encounter by
$f_e$ we have from the equation of path (cf. \cite{curtis2}) that
\be
r_e=\frac{a(1-e^2)}{1+e \cos f_e}
\ee
where $r_e$ is the radius of Earth's orbit, $a$ is the semimajor axis
of the asteroid's orbit, and $e$ is the eccentricity of the asteroid's
orbit. Thus
\be
f_e=\cos^{-1}\left(\frac{1}{e}(\frac{a(1-e^2}{r_e}-1)\right)
\ee
or
\be
f_e=-1.3 \mbox{ rad}
\ee

The deflection of the asteroid for various parameter values will be calculated using (\ref{sums}). Thus we start by calculating the various quantities that are needed in that equation.

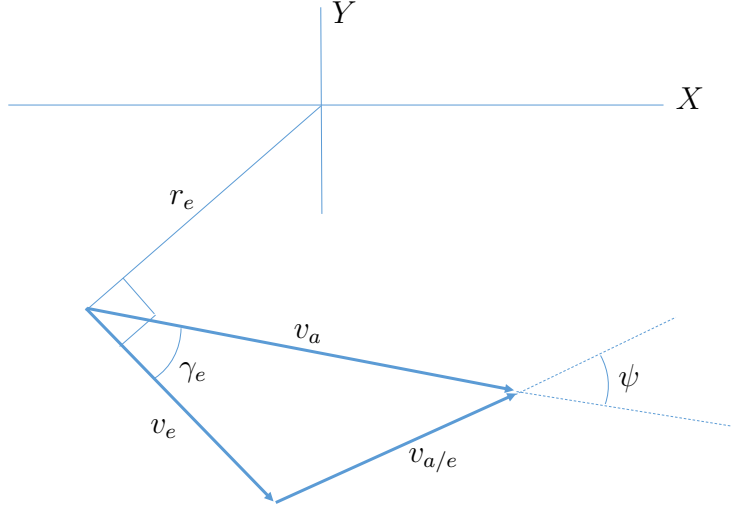
\begin{figure}[htbp]
  \begin{center}
    \unitlength=.5in
    \begin{picture}(4,4.6)
    \put(-4,-1.7){\includegraphics[scale=.5]{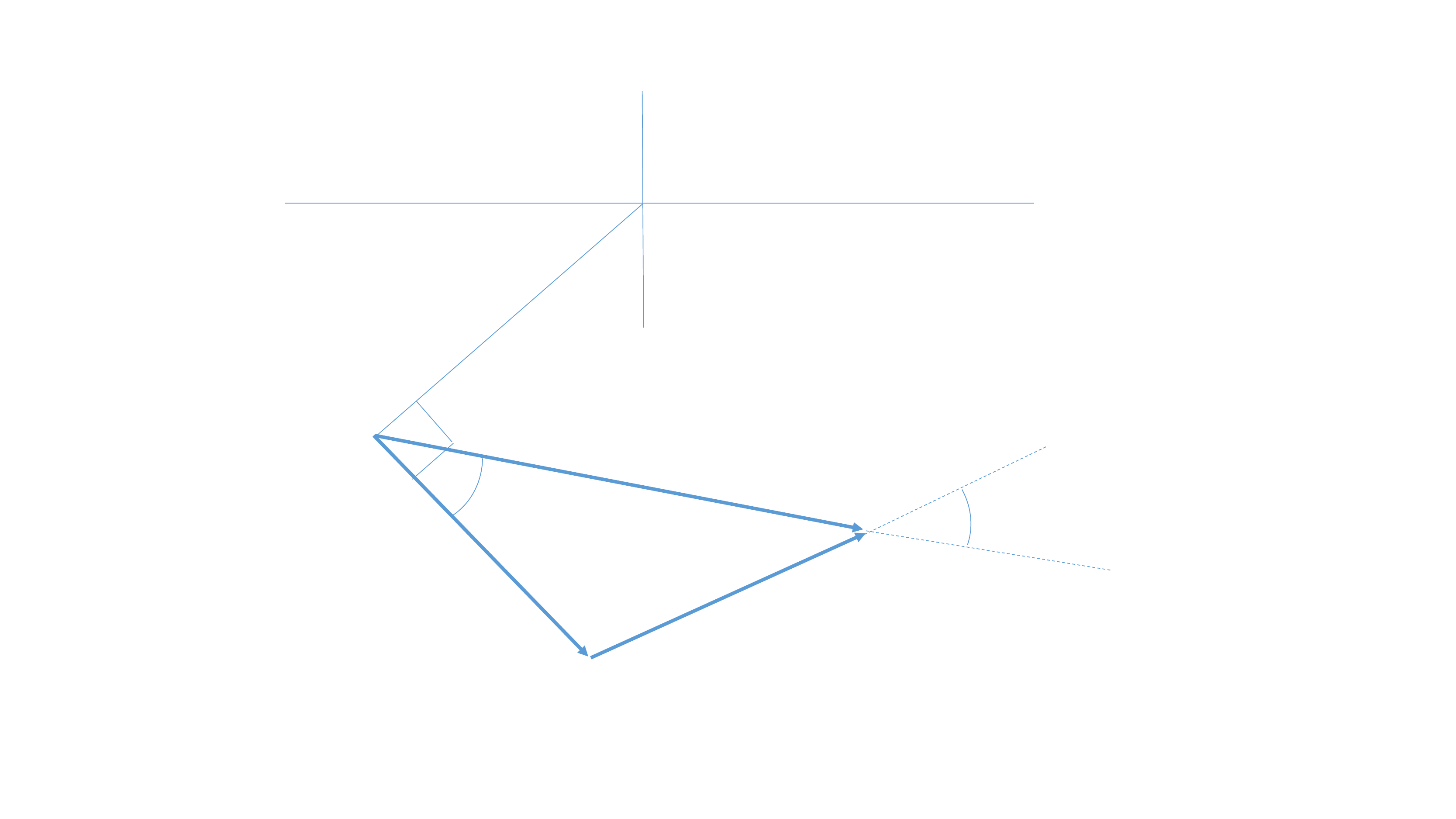}}
     \put(.4,1.1){\large$\gamma_e$}
     \put(5,1){\large$\psi$}
     \put(.3,2.9){\large$r_e$}
     \put(.1,.5){\large$v_e$}
     \put(1.6,1.5){\large$v_a$}
     \put(2.8,0.2){\large$v_{a/e}$}
     \put(5.6,3.9){\large$X$}
     \put(2,4.8){\large$Y$}
    \end{picture}
  \end{center}
  \caption{Geometry of the encounter of Earth and VK184 in June 2048.}
\label{geometry}
\end{figure}

The velocity of the asteroid at encounter is obtained from the energy
equation
\be
v_{ae}=\sqrt{2\left(\frac{\mu_s}{r_E}-\frac{\mu_s}{2a_a}\right)}
\ee
or
\be
v_{ae}=35500 \mbox{ m/s}
\ee
which from (\ref{kappa}) gives for $\kappa$
\be
\kappa=1.528\times10^{-4} \mbox{ s/m}
\ee
to be used in (\ref{sumsq2}).
The flight path angle $\gamma$ (cf. \cite{curtis2}) for the asteroid at the time of encounter can
now be calculated from the 
expression for the angular momentum, (cf. \cite{curtis2})
\be
h=rv_{ae}\cos\gamma=\sqrt{a(1-e^2)\mu_s}
\ee
or
\be
\gamma=\cos^{-1}\frac{\sqrt{a(1-e^2)\mu_s}}{rv_{ae}}
\ee
or
\be
\gamma=0.44 \mbox{ rad}
\ee
Next referring to Figure \ref{geometry} we calculate the magnitude of
the relative velocity of the asteroid with respect to Earth as
\be
v_{a/E}=\sqrt{v_E^2+v_a^2-2v_av_E\cos\gamma}
\ee
or
\be
v_{a/E}=15,200 \mbox{ m/s}
\ee
It then follows for the incidence angle $\psi$ that
\be
\frac{v_{a/E}}{\sin\gamma}=\frac{v_E}{\sin\psi}
\ee
or after solving for $\psi$
\be
\psi=0.829 \mbox{rad}
\ee

For the gravity tractor, we assume a gross mass of $m_{c1}=1500$ kg and a fuel
mass of $m_f=450$ kg. (As will be seen below, this corresponds to a
mission duration of about 6 years). Following \cite{olympio} the
specific impulse of the spacecraft will be  assumed to be $I_{sp}=2500$ s.  

\begin{figure}[htbp]
  \begin{center}
    \unitlength=.5in
    \begin{picture}(4,5)
    \put(-2.15,-.2){\includegraphics[]{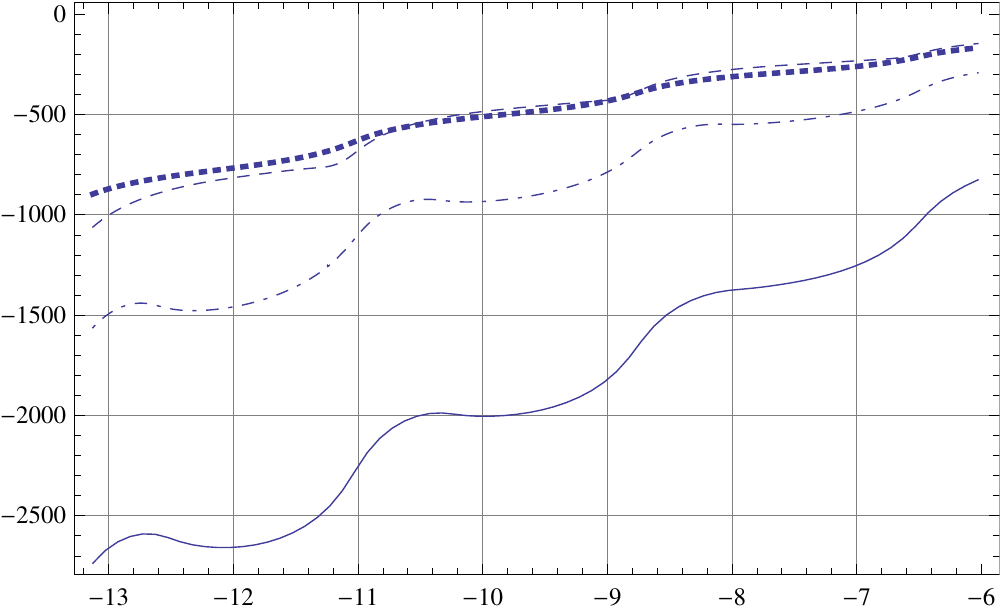}}
      \put(6,-.3){$t_s-t_e$ (years)}
      \put(-3,4.5){$\Delta\zeta$ (km)}
    \end{picture}
  \end{center}
  \caption{Comparison of deflections obtained by using the Keplerian gravity tractor (solid line); the stationary gravity tractor: $\alpha=1.5$: (dashed line), $\alpha=2.5$: (dashed-dotted line) as functions of the start time before the projected encounter.}
\label{vk184}
\end{figure}

From Figure \ref{times} we pick $\theta_b=1$ rad for a time of flight
of around $1$ TU.  From Figure \ref{conics} we pick for simplicity $e=0$ for
$\eta_k\approx 0.8$. Using (\ref{rpm}) we calculate the smallest
radius that satisfies the plume impingement constraint to be
$r_{pm}= 69.2$ m. Using (\ref{tcirc}) the time of flight on the orbit
segment is then calculated to be
\be
\Delta t=2452 {\mbox{s}},
\ee
and using (\ref{dv}) the required $\Delta v$ at the ends of the
segment are calculated to be
\be
\Delta v=0.1128 {\mbox{m/s}}
\ee
This value of $\Delta v$ used in (\ref{q1}) gives
\be
q=4.602\times10^{-6}
\ee
The value of $N$ is calculated based on how many times the impulsive
thrusts can be imparted before all the fuel is expended, or
\be
m_{c1}-m_f=m_{c1}e^{-qN}
\ee
or
\be
N=-\frac{1}{q}\ln{\left(\frac{m_{c1}-m_f}{m_{c1}}\right )}
\ee
which yields
\be
N=77,510
\ee
The total duration of a mission $t_m$ is therefore
\be
t_m=N\Delta t
\ee
or $t_m=6.03$ years.
Next using (\ref{beta}) we find 
\be
\lambda=3.87\times10^{-5} \mbox{ m/s}^2
\ee
Lastly we note that the velocity $v_{ai}$ needs to be determined
for each value of the index $i$. Thus for a given $i$ we start by solving Kepler's equation for the corresponding eccentric anomaly $E_i$ using the known eccentric anomaly at the time of encounter $E_e$ 
\be
t_i-t_e=\sqrt{\frac{a^3}{\mu_s}}(E_i-E_e-e(\sin(E_i)-\sin(E_e)))
\label{kepler}
\ee
Using $E_i$ we calculate the corresponding radius $r_i$ from the equation of path (see \cite{curtis3})
\be
r_i=a_a(1-e \cos E_i)
\label{rai}
\ee
which can be used in the energy equation to yield
\be
v_{ai}=\sqrt{2\left(\frac{\mu_s}{r_i}-\frac{\mu_s}{2a_a}\right)}
\label{vai}
\ee
The above numerical values can now be used in (\ref{sums}) for a range of
values of the starting time before the time of projected encounter $t_s-t_e$. For each value of $i$ in (\ref{sums}) the corresponding value of $v_{ai}$ is determined through a numerical solution of Kepler's equation (see to yield the total deflection as a
function $t_s-t_e$. The result is depicted in Figure \ref{vk184}
(solid line). 

We next consider a spacecraft of the same gross and fuel masses as
above, this time deployed as a stationary gravity tractor. As the mass of the
spacecraft decreases the thrust decreases in order for the 
spacecraft to remain at equilibrium in its stationary position. While
other scenarios are 
possible (such as a decreasing altitude for a constant gravitational force) this is the most efficient approach as discussed in
\cite{sanchez2}.

Thus it follows from (\ref{FT}) that the
thrust required to keep the spacecraft at equilibrium also decreases
according to the relation
 
\be
T=\frac{Gm_am_{c}}{2\alpha^2 r_a^2\cos(\beta_0+\varphi)}
\label{thrustreq}
\ee
Now using (\ref{thrustreq}) in (\ref{tisp}) and recalling $2dm'=dm_c$ gives
\be
\frac{dm_c}{dt}=-\frac{2T}{I_{sp}g_0}
\label{dmdt}
\ee
or
\be
\frac{dm_c}{dt}=-\frac{Gm_am_c}{I_{sp}g_0\alpha^2
r_a^2\cos(\beta_0+\varphi)}
\ee
Solving this differential equation we have
\be
m_c=m_{c1}e^{-Q(t-ts)}
\label{mcoft}
\ee
where we have defined the parameter $Q$ through
\be
Q=\frac{Gm_a}{I_{sp}g_0\alpha^2
r_a^2\cos(\beta_0+\varphi)}
\label{Q2}
\ee
and where $m_{c1}$ is the mass of the spacecraft at the initial time
$t_s$, i.e. the same initial mass as for the Keplerian gravity tractor.

To obtain the deflection caused by the stationary gravity tractor we can directly use (\ref{constf}):
\be
\Delta \zeta^c=-\kappa\int_{t_s}^{t_f}(t_e-t)v_a\frac{Gm_{c1}e^{-Q(t-t_s)}}{\alpha^2 r_a^2}dt
\label{stat}
\ee
We start by calculating $Q$ using
(\ref{Q2}) which yields
\be
Q=4.50\times 10^{-9}\mbox{ /s}
\ee
Next from (\ref{mcoft}) we find that the longest the stationary
gravity tractor can operate is
\be
t_f-t_s=-\frac{1}{Q} \ln {\left( \frac{m_{c1}-m_f}{m_{c1}} \right )}
\ee
or
\be
t_f-t_s=2.51 \mbox{ years}
\ee
The integral in (\ref{stat}) is now evaluated numerically taking into 
account the variation of $v_a$ with time according to (\ref{kepler})-(\ref{rai}).
Thus we calculate the deflection created at the
time of encounter for a range of values of the starting time before
the projected encounter
$t_s-t_e$. The results using $\alpha=1.5$ and $\alpha=2.5$ are shown in Figure \ref{vk184}: dashed line and dashed-dotted line, respectively.

Lastly, we determine the deflection created by a displaced-orbit gravity tractor of the same initial mass. Using (\ref{tee}) (with $m'=m$ as there is only one thruster) and (\ref{fd})
\be
\frac{dm_c}{dt}=-0.21\frac{Gm_am_c}{I_{sp}g_0}
\ee
or
\be
m_c=m_{c1}e^{-Q_d(t-ts)}
\label{mcoftd}
\ee
where
\be
Q_d=0.21\frac{Gm_a}{I_{sp}g_0}
\label{Q3}
\ee
The value of $Q_d$ is obtained to be
\be
Q_d=4.46\times10^{-10}/s
\ee
Next, from (\ref{mcoftd}), the longest the gravity tractor can operate is found to be
\be
t_f-t_s=-\frac{1}{Q_d} \ln {\left( \frac{m_{c1}-m_f}{m_{c1}} \right )}
\ee
or
\be
t_f-t_s=25.3 {\mbox{years}}
\ee

Next, using we use (\ref{dzetatot}) we note that the total deflection
at time $t_e$ may be written as
\be
\Delta \zeta^d=-0.21 \kappa
\int_{t_s}^{t_f}(t_e-t)v_a\frac{Gm_{c1}e^{-Q_d(t-t_s)}}{r_a^2}dt
\label{disp}
\ee
The integral in (\ref{disp}) is evaluated numerically
for a range of values of the starting time before
the projected encounter
$t_s-t_e$. The result is shown in Figure \ref{vk184} (dotted line).

In summary it is clear that the Keplerian gravity tractor can result
in a deflection that is significantly larger than that for a stationary gravity
tractor for a given lead time and initial mass of the spacecraft.
The added deflection ranges from about 1000 km for a deployment 6
years ahead of the projected time of encounter, to more than 1500 km for a
deployment 12 years ahead of the time of encounter.
Recalling the discussion in the previous section, this result
may be explained by the larger average force that can be exerted by
the Keplerian gravity tractor as well as its overall higher fuel
efficiency.

Another interpretation of Figure \ref{vk184} is in terms of what lead
time is necessary for a given deflection. Thus we note for example
that in order to realize a deflection of 1000 km, the stationary
gravity tractor would need a 13-year lead time, about twice what
would be required for the Keplerian gravity tractor (about 6.5 years).

The result in Figure \ref{vk184} may also be compared to a similar
result in \cite{olympio} where an example of an optimized stationary
gravity tractor with slightly larger mass is discussed.
It is seen that the deflections resulting
from the Keplerian gravity tractor are of the same order of magnitude
as that of the optimized stationary gravity tractor, despite the
smaller initial mass of the spacecraft.

\section{Conclusions}

While the probability of a major asteroid impact is small, its potential for
severe consequences on Earth makes it worthwhile to study possible
methods of deflection or impact mitigation. Asteroid deflection
methods based on gravitational coupling constitute a promising 
approach for averting a collision. A central challenge associated
with these methods is that the magnitude of the force that can be
exerted using a spacecraft is generally small which means that a
significant deflection requires a large lead time. 

The asteroid deflection scheme based on
Keplerian orbits described in this paper
addresses this issue and describes a method for a more efficient
gravity tractor in terms of an increased force on the asteroid
for a given mass of a spacecraft. In addition better performance is
attained with regards to fuel consumption. Further, the dynamic nature of
the gravity tractor makes it possible for the 
simultaneous deployment of several spacecraft which could decrease the
time required for a given deflection.

Additional enhancement of the efficiency of the Keplerian gravity tractor
can be expected through an optimization of its operation. As in
\cite{olympio} this would mean that the spacecraft would
be in a  ``tugging'' mode,  i.e. in restricted Keplerian motion, only
when the velocity of the asteroid is sufficiently
large to warrant the fuel expenditure. At other times the spacecraft
would simply orbit the asteroid without imparting a net impulse over
time. Thus the optimized gravity tractor is especially suited for asteroids on highly
elliptic orbits where a large variation in asteroid velocity occurs.
The details of this approach are the subject of current work.


\end{document}